\documentclass[useAMS,usenatbib]{mnras}

\usepackage{graphicx}
\usepackage{multirow}
\usepackage{tabularx}  
\usepackage{caption}
\usepackage{subcaption}
\usepackage{changes} 
\usepackage{ulem}

\newcommand{\Msol}{M$_\odot$}
\newcommand{\rhocrit}{$\rho_\textrm{\small crit}$}
\newcommand{\OmegaB}{$\Omega_\textrm{\small b}$}
\newcommand{\OmegaDM}{$\Omega_\textrm{\small dm}$}

\title[Large-scale mass distribution]{Large-scale mass distribution in the 
Illustris simulation}
\author[M. Haider et al.]{M. Haider$^{1}$\thanks{E-mail:
markus.haider@uibk.ac.at}, D. Steinhauser$^{1}$, M. Vogelsberger$^{2}$, S. Genel$^{3}$\thanks{Hubble fellow}, 
V. Springel$^{4,5}$\newauthor P. Torrey$^{2,6}$ and L. Hernquist$^{7}$\\
$^{1}$Institut f\"ur Astro- und Teilchenphysik, Universit\"at Innsbruck, Technikerstra\ss e 25/8, A-6020 
Innsbruck, Austria\\
$^{2}$Kavli Institute for Astrophysics and Space Research, Massachusetts Institute of Technology, Cambridge, 
MA 02139, USA\\
$^{3}$Department of Astronomy, Columbia University, 550 West 120th Street, New York, NY 10027, USA\\
$^{4}$Heidelberg Institute for Theoretical Studies, Schloss-Wolfsbrunnenweg 35, D-69118 Heidelberg, Germany\\
$^{5}$Zentrum f\"ur Astronomie der Universit\"at Heidelberg,
  Astronomisches Recheninstitut, M\"{o}nchhofstr. 12-14, D-69120
  Heidelberg, Germany\\
$^{6}$TAPIR, California Institute of Technology, Mailcode 350-17, Pasadena, CA 91125, USA\\
$^{7}$Harvard-Smithsonian Center for Astrophysics, 60 Garden Street, Cambridge, MA 02138, USA
}

\begin{document}

\date{Accepted 2016 January 8. Received 2015 December 21; in original form 2015 August 6}

\pagerange{\pageref{firstpage}--\pageref{lastpage}} \pubyear{2015}

\maketitle

\label{firstpage}

\begin{abstract} 
  Observations at low redshifts thus far fail to account for all of the baryons expected in the
  Universe according to cosmological constraints. A large fraction of the baryons presumably
  resides in a thin and warm--hot medium between the galaxies, where they are difficult to observe
  due to their low densities and high temperatures. Cosmological simulations of structure formation
  can be used to verify this picture and provide quantitative predictions for the distribution of
  mass in different large-scale structure components. Here we study the distribution of baryons
  and dark matter at different epochs using data from the Illustris simulation. We identify 
  regions of different dark matter density with the primary constituents of large-scale
  structure, allowing us to measure mass and volume of haloes, filaments and voids. At redshift zero,
  we find that \mbox{49 \%} of the dark matter and \mbox{23 \%} of the baryons are within haloes
  more massive than the resolution limit of $2\times 10^8$ \Msol. 
  The filaments of the cosmic web host a further \mbox{45 \%} of the dark matter and \mbox{46 \%}
  of the baryons. The remaining \mbox{31 \%} of the baryons reside in 
   voids. The majority of these baryons have been transported there through
  active galactic nuclei feedback. We note that the feedback model of Illustris is too strong for heavy haloes, therefore
  it is likely that we are overestimating this amount.
  Categorizing the baryons according to their density and temperature, we find that \mbox{17.8 \%} of
  them are in a condensed state, \mbox{21.6 \%} are present as cold, diffuse gas, and \mbox{53.9 \%} are
  found in the state of a warm--hot intergalactic medium.  
\end{abstract}

\begin{keywords}
galaxies: haloes -- cosmology: dark matter -- large-scale structure of Universe.
\end{keywords}

\section{Introduction}

The large-scale structure of the Universe is determined by the dominant dark matter, which makes up \mbox{26 
\%} of today's mass--energy content of the Universe. As dark matter is not detectable directly, the 
large-scale structure must be inferred from the baryons. Recent analysis \citep{Collaboration2015} of the 
cosmic microwave background fluctuations by the \textit{Planck} mission found that baryons amount to \OmegaB\ 
= \mbox{4.8 \%} of the present-day critical density of the Universe. This corresponds to a mean baryon 
density of $\rho_\textrm{b} = 4.1\times 10^{-31} \textrm{g}\,\textrm{cm}^{-3}$. An independent measurement of 
the baryon content of the Universe is possible through the theory of big bang nucleosynthesis, as it allows 
one to infer the baryon content from the abundance ratio of light elements. The measurements of the abundance 
ratio of deuterium to hydrogen (D/H) \citep{Kirkman2003} are in very good agreement with the baryon content 
derived from the cosmic microwave background fluctuations. 
At higher redshifts, the baryons were not fully ionized. This allows their detection in denser regions 
through \textsc{Hi} absorption lines in the spectrum of background quasars, the so-called Lyman alpha 
(Ly\,$\alpha$) forest. The amount of gas needed to produce the observed absorption lines in the Ly\,$\alpha$ 
forest is in good agreement with the cosmic baryon fraction \OmegaB\ \citep{Weinberg1997}.

At low redshifts, however, observations fail to account for all the baryons detected at higher redshifts 
\citep{Fukugita2004,Bregman2007}. While X-ray observations of galaxy clusters do find baryon contents close 
to the primordial value \citep{Simionescu2011}, and \textsc{Ovii} absorption line measurements at X-ray 
energies hint at large reservoirs of hot gas around spiral galaxies \citep{Gupta2012}, approximately \mbox{30 
\%} of the baryons are still not accounted for in the local Universe \citep{Shull2012a}. The baryons are not 
only missing on cosmological scales, but the baryon-to-dark matter ratio in galaxies also falls short of the 
primordial ratio \citep{Bell2003,McGaugh2009}.

Based on hydrodynamical simulations \citep{Dave2000}, we expect that a significant amount of the baryons are
hidden in the state of a thin warm--hot intergalactic medium (WHIM) with temperatures between 10$^5$ and 
10$^7$ K. Through \textsc{H\,i} and \textsc{O\,vi} surveys \citep{Danforth2008}, it is possible to probe the 
colder parts of the low-redshift intergalactic medium. However, the gas is too diffuse and the temperatures 
are insufficient to be detectable with the current generation of X-ray satellites \citep{Kaastra2013}, except 
in the densest regions between galaxy clusters \citep{Nicastro2005b,Nicastro2013}. The question of how the 
baryons are distributed and whether they follow the filaments of the cosmic web is not well constrained 
observationally, and consequently, cosmological simulations of structure formation are important tools for 
making predictions about the large-scale distribution of baryons 
\citep{Cen1999,Cen2005,Dave1999,Dave2000,Smith2010}.
In this paper, we investigate the distribution of baryons and dark matter using data from the Illustris 
simulation \citep{Vogelsberger2014}. In \autoref{sec:simulation}, we give a short overview of the simulation 
and discuss the methods we used. Our results are presented in \autoref{sec:results}. Specifically, in 
\ref{sec:halos}, we investigate the amount of matter inside haloes and compute the baryon fraction for haloes 
of different masses. In \autoref{sec:dm_density}, we examine the distribution of matter with respect to the 
dark matter density. This allows us to decompose the simulation volume into haloes, filaments and voids, and 
measure the mass, volume and redshift evolution of these components. In \autoref{sec:whim}, we look at the 
distribution of the baryons in the temperature--density space and compute the mass fraction and redshift 
evolution of the condensed, diffuse, hot and WHIM phases. We discuss the
results in \autoref{sec:discussion} and close with a summary in \autoref{sec:summary}.

\section{Simulation \& Methods}\label{sec:simulation}
We analyse data from the Illustris simulation, which is a cosmological hydrodynamics simulation of galaxy 
formation in a (106.5 Mpc)$^3$ volume [corresponding to a (75 Mpc/h)$^3$ periodic box]. The gas and dark 
matter were evolved using the \textsc{arepo} moving-mesh code \citep{Springel2010a}. In addition to gravity 
and hydrodynamics, the simulation includes subgrid models for star formation, active galactic nuclei (AGN) feedback and gas cooling 
[see \citet{Vogelsberger2013} and \citet{Torrey2014} for details on the implemented feedback models and the 
parameter selection]. Dark matter and gas are represented by 1820$^3$ resolution elements each, resulting in 
an initial mass resolution of 1.26$\times 10^6 $ \Msol\ for gas and 6.26$\times 10^6 $ \Msol\ for dark 
matter. The simulation uses a standard $\Lambda$CDM model with $H = 70.4$, $\Omega_0 = 0.2726$, 
$\Omega_\Lambda = 0.7274$ and \OmegaB = 0.0456. It was started at redshift $z = 127$ and evolved all the way 
to the present epoch. The Illustris simulation produced a galaxy mass 
function and a star formation history which are in reasonably good agreement with observations. Remarkably, 
it also yielded a realistic morphological mix of galaxies. An overview of basic results is given in 
\citet{Vogelsberger2014}, \citet{Vogelsberger2014a} and \citet{Genel2014}, and recently, all of the 
simulation data has been made publicly available \citep{Nelson2014}.

In \autoref{sec:halos}, we use the halo catalogue of the Illustris simulation, which has been generated using 
the \textsc{subfind} halo-finding algorithm \citep{Springel2000}. \textsc{subfind} searches for 
gravitationally bound overdensities and computes the dark matter, gas and stellar mass which is bound to a 
halo or subhalo. In the subsequent analysis, we compute the density fields of gas, stars, baryonic matter and 
dark matter on a uniform grid with 1024$^3$ cells. To this end, the dark matter and gas particles are mapped 
conservatively to the grid using a smoothed particle hydrodynamics (SPH) kernel technique. For the dark matter, we set the smoothing length to 
be the radius of a sphere containing the 64 nearest dark matter particles, as is done also in 
\textsc{subfind}. The gas cells are mapped using three times $r_\textrm{c}=(3V_\textrm{\small cell}/{4\pi})^{1/3}$, 
where $V_\textrm{\small cell}$ is the volume of a Voronoi cell. The stellar and black hole particles have 
been mapped using the particle in cell method with a nearest grid point 
assignment, as they are not tracers of an extended density field. We should note that the densities given in 
this analysis are thus average densities over the volume of individual cells, which are 104 kpc on a side.

The baryon densities and temperatures used in \autoref{sec:whim} have been computed directly from the Voronoi 
cells of the simulation and do not involve any additional smoothing. 
For comparison, we also use a variant of the Illustris simulation where cooling, star formation and feedback 
have been switched off. This simulation, which we will refer to as non-radiative run, has only half the 
resolution of the full physics simulation.

\section{Results}\label{sec:results}

In \autoref{sec:halos}, we investigate the mass fraction of baryons and dark matter in haloes and compute the 
baryon-to-halo mass ratio. In \autoref{sec:dm_density}, we examine the mass distribution with respect to the 
dark matter density. With the help of cuts in dark matter density, we divide the simulation volume into 
haloes, filaments and voids. In \autoref{sec:whim}, we analyse the baryon distribution in temperature--density 
space.

\subsection{Mass in haloes}\label{sec:halos}

In \autoref{table:mass_in_halos}, we give the mass fraction of dark matter and baryons residing inside the 
identified \textsc{subfind} haloes. The mass fractions are calculated using the mass of all particles which 
are gravitationally bound to haloes. Only main haloes were considered, with the mass of genuine subhaloes 
being added to their parent halo. The table also shows the decomposition of the baryonic mass into gas and 
stars. We should note that the mass found in haloes depends on the resolution of a simulation, therefore 
\autoref{table:mass_in_halos} only gives the fraction of mass in haloes which can be resolved in the 
Illustris simulation. The resolution limit in terms of mass is primarily given by the mass of a single dark 
matter particle which is $6.26\times 10^6$ \Msol. To detect gravitationally bound haloes, a minimum of about 
30 particles are required. Thus, the threshold for halo detection in Illustris is 
approximately $2\times 10^8$ \Msol. A resolution study using a run with one-eighth of the mass resolution leads to similar 
results as the ones presented in \autoref{table:mass_in_halos}. This suggests that the results
are numerically robust. How much of the dark matter is expected to reside 
in haloes below the resolution limit also depends on the physical nature of the dark 
matter particle. \citet{Angulo2010} find that for a 100 GeV neutralino dark matter particle only 5--10 \% of 
the dark matter should not be part of a clump at redshift zero, and that the most abundant haloes 
have masses around $10^{-6}$ \Msol.

We find that about half of the total dark matter mass in the simulation volume is contained within haloes, 
while only \mbox{21 \%} of the baryonic mass resides within haloes. This evidently means that in the 
Illustris simulation, the mass ratio between baryons and dark matter is on average lower in haloes than the 
primordial mixture, \OmegaB/\OmegaDM. The latter would be the expected value if baryons traced dark matter 
perfectly. The second row of \autoref{table:mass_in_halos} shows that nearly all the halo baryons are in 
haloes with a total mass higher than 5$\times 10^9$ \Msol. Less massive haloes in Illustris consist mainly of 
dark matter and are hence truly dark. We should note though that a $10^9$ \Msol\ halo is made up of only 160 
dark matter particles, and the results will be less accurate at this scale than for haloes with higher 
numbers of particles (\citet{Vogelsberger2014a}, see \citet{Trenti2010a} for a discussion of the influence of particle number
on the quality of halo properties). For haloes more massive than 5$\times 10^9$ \Msol, we again 
see that the baryon/dark matter ratio is only half of the primordial ratio, and in the mass 
range between $10^{12}$ and $10^{13}$ \Msol, it drops to about one-third. Towards the most massive haloes of 
cluster size, this rises again to about \mbox{50 \%} and beyond of the expected primordial value.

Comparing these mass fractions bound in haloes with the results of the non-radiative run of the Illustris 
initial conditions, we find that the amount of dark matter in haloes is very similar. However, in the 
non-radiative run, the fraction of baryons in haloes is much higher and reaches about \mbox{40 \%}, which is 
roughly double the value of the full physics run. As the overall dark matter structure is very similar for 
the full physics and the non-radiative run, the difference in baryonic mass is a consequence of the 
feedback processes included in the full physics simulation (see also Sections \ref{sec:dm_density} and 
\ref{sec:discussion}).

The mass fraction in haloes also depends on the definition of halo mass. The values presented in 
\autoref{table:mass_in_halos} are the total gravitationally bound masses. Another common way to define the 
mass of haloes is to count the mass inside a sphere whose radius is set so that the mean enclosed density 
equals a reference density. If we measure the mass around the primary haloes of the  \textsc{subfind} 
friend-of-friend groups, and use as a reference density $200\times\rho_\textrm{\small mean}$ (where 
$\rho_\textrm{\small mean}$ denotes the mean matter density of the Universe) we find a total mass fraction of 
48.1 \% in haloes. Using the higher reference density $500\times\rho_\textrm{\small crit}$ this value reduces 
to 26 \%. We also want to note that for high-mass haloes, it makes a difference if the mass includes the 
contribution of subhaloes or not. In Illustris, 19.2 \% of the haloes are subhaloes. They make up for 13.7 \% 
of the total mass and 29.7 \% of the stellar mass. Thus, if we neglected the 
contribution of subhaloes, the haloes with a mass higher than $10^{14} $\Msol\ would be 
found to host only 4.4 \% of the total matter, 4.9 \% of the dark matter and 1.9 \% of the baryons.

\begin{table*}
\caption{Mass fraction of dark matter and baryons inside haloes at $z=0$. The mass fractions are given with 
respect to the total mass inside the simulation volume. The first row shows the fractions for all haloes in 
the Illustris full physics simulation. The second row gives the mass contribution of haloes with a total mass 
higher than $5\times 10^9$ \Msol\ and the third and fourth rows show the respective values for haloes more 
massive than $10^{12}$ and $10^{13}$ \Msol, respectively. The last row gives the mass fraction for all 
haloes of the non-radiative Illustris run (no star formation, feedback or cooling).}
\label{table:mass_in_halos}
\centering
\begin{tabular}{lccccc}
\hline
& \% of total   & \multicolumn{3}{c}{\% of total baryonic mass} & \% of total\\
& dark matter mass & baryons & gas & stars & mass\\ \hline
all haloes & 50.4 \% & 21.3 \% & 14.7 \% & 6.6 \%& 45.5 \% \\
M$_\textrm{tot} >$ 5$\times 10^9$\Msol & 44.6 \% & 21.3 \% & 14.7 \% & 6.6 \%& 40.7 \% \\
M$_\textrm{tot} > 10^{12}$\Msol & 29.3 \% & 9.8 \% & 4.8 \% & 5.0 \%& 26.0 \% \\
M$_\textrm{tot} > 10^{13}$\Msol & 20.1 \% & 7.2 \% & 4.0 \% & 3.2 \%& 18.0 \% \\
M$_\textrm{tot} > 10^{14}$\Msol & 11.8 \% & 6.0 \% & 3.7 \% & 2.3 \%& 10.8 \% \\
all haloes, non-radiative run & 49.8 \% & \multicolumn{2}{c}{39.1 \%} & -& 48.0 \% \\
\hline
\end{tabular}
\end{table*}

\begin{figure}
\includegraphics[width=0.49\textwidth]{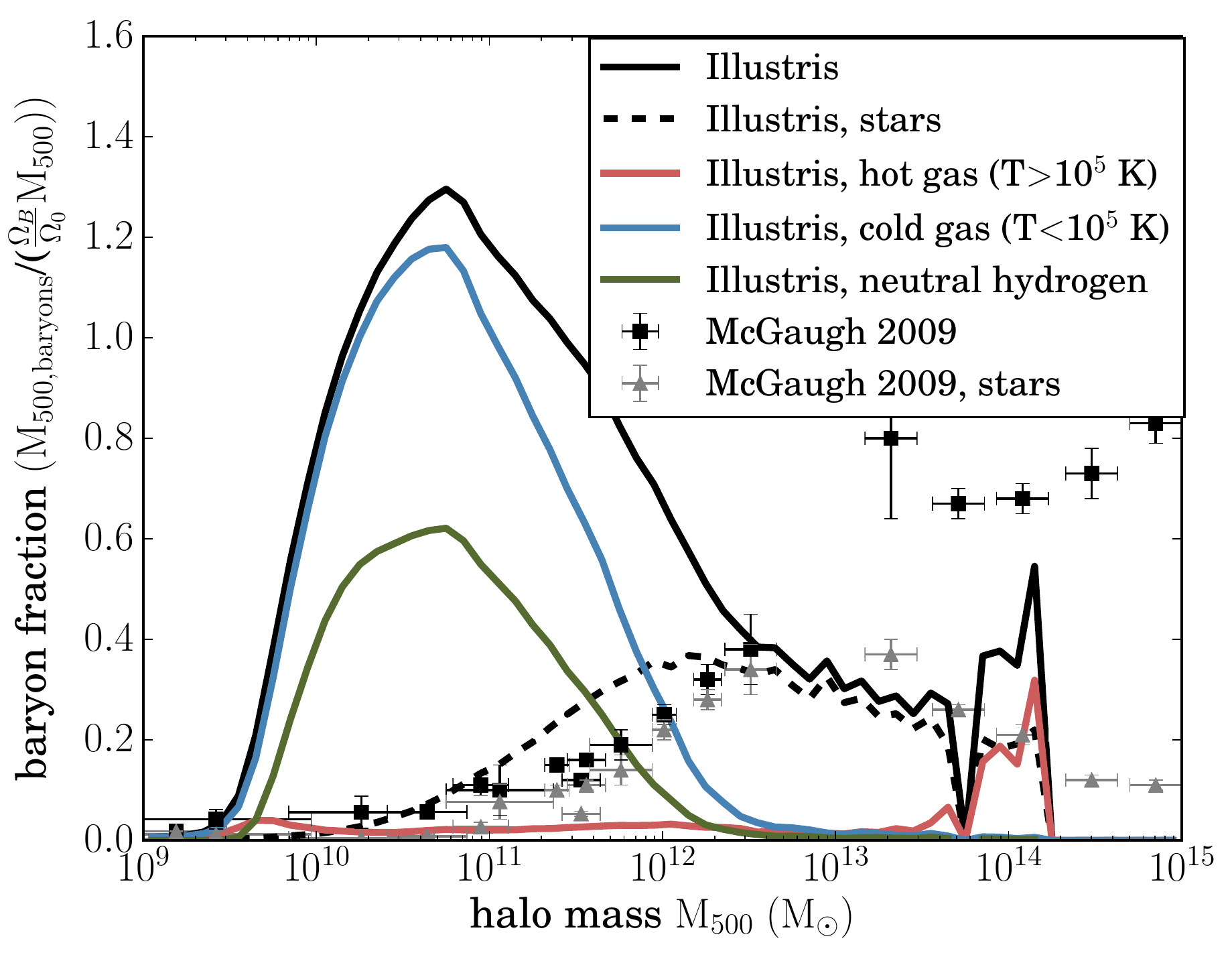}
\caption{Baryon to total matter ratio inside central haloes of the Illustris simulation divided by 
the primordial ratio \OmegaB/$\Omega_\textrm{\small 0}$ (solid black line). The $x$-axis gives the total mass 
of haloes inside a sphere with a mean density of 500 times the critical density. The dashed black line shows the 
contribution of stars, the red line the contribution of gas warmer than \mbox{10$^5$ K}, the blue line gas 
colder than \mbox{10$^5$ K} and the green line the amount of neutral hydrogen. The black squares and grey 
triangles represent observational data, taken from table 2 of \citet{McGaugh2009}.}
\label{fig:baryon_fraction_halos}
\end{figure}

We further illustrate the baryon-to-dark matter ratio in \autoref{fig:baryon_fraction_halos}, where we plot 
the ratio of the baryon content M$_\textrm{\small baryon}/$M$_\textrm{\small total}$ to the primordial baryon 
fraction \OmegaB/$\Omega_\textrm{\small 0}$ against the halo mass. The baryon fraction is broken down into components of cold 
gas, hot gas, neutral hydrogen and stars, and the plot includes observational data compiled by 
\citet{McGaugh2009} for comparison.
\begin{figure}
\includegraphics[width=0.45\textwidth]{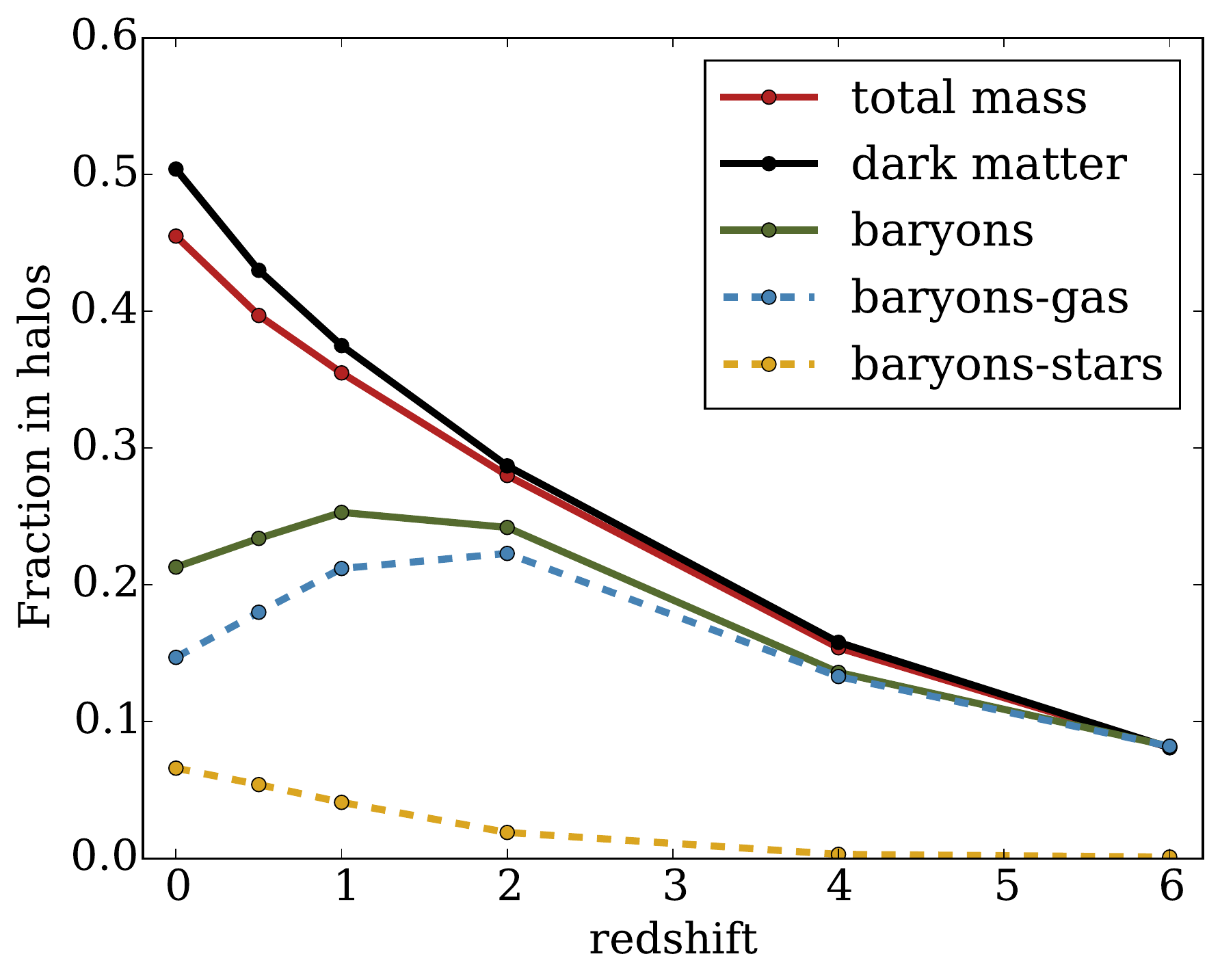}
\caption{Redshift evolution of the mass inside haloes. The values for dark matter and baryons are normalized 
to the total dark matter and baryon mass in the simulation volume, respectively. The dashed lines show the 
contribution of gas and stars to the baryon fraction.}
\label{fig:mass_halos_evo}
\end{figure}
Only central haloes (those which are not subhaloes) have been used for this plot and only the mass inside 
$R_\textrm{\small 500c}$ (the radius inside of which the mean density is 500 times the critical density) is 
taken into account. A resolution study using data from a run with one-eighth of the mass resolution produced results
very similar to \autoref{fig:baryon_fraction_halos}.

In the regime of galaxy clusters, X-ray observations find a 
baryon to dark matter ratio between 70 and 100 \% of the primordial value \citep{Vikhlinin2005}. For cluster 
sized haloes, Illustris finds a ratio around 40--50 \%, but only small clusters are present in the limited 
simulations volume. This disagreement with observations has already been noted \mbox{\citep[see Fig. 10 in][]{Genel2014}} and 
has been attributed to a too violent radio-mode AGN feedback model. The radio-mode feedback model is used when the black hole accretion rates are low. It is most 
effective at removing gas from the haloes in the mass range from $10^{12.5}$ to $10^{13.5}$\Msol, as at lower
halo masses, the black holes are less massive and at higher halo masses, the potential wells 
are deeper. The model and the used parameters result in too large mass outflows from group-sized haloes and poor clusters (see \autoref{sec:discussion} for a 
more detailed discussion of the model). 
As a considerable amount of mass is within group-sized haloes and clusters, the results about the distribution of baryons, 
which we present in the following sections, will be affected by the mass outflows.

Between $10^{10}$ \Msol\ and $10^{11}$ \Msol, the baryon fraction is
close to the primordial value \OmegaB/$\Omega_\textrm{\small 0}$. In this mass regime, the
radio-mode AGN feedback plays only minor role as the black holes are
less massive. The applied supernova feedback model is removing gas
from the star-forming phase through galactic winds. However, the
velocity of these winds is lower than the escape velocity and thus the
supernova feedback model does not remove gas from the
haloes. We find that most of the baryons in this mass range are
  at relatively low temperatures, and a large fraction of the gas is
  in neutral hydrogen.  We should note though, that Illustris does not
  simulate the effects of the stellar radiation field on the gas, and
  therefore we cannot predict the ionization state exactly.  The
  baryon fraction we find is higher than the fraction of detected
  baryons in \citet{McGaugh2009}. However, these observations
  should be thought of as lower limits. A large fraction of the
  baryonic content of galaxies is expected to reside within the
  circumgalactic medium, which is subject to large observational
  uncertainties and is depending on model assumptions. Recent studies
  \citep{Tumlinson2013,Bordoloi2014,Werk2014} suggest that galaxies
  can have substantial amounts of cold gas in their circum-galactic
  medium. In particular, we note that the baryon fraction we find in
  Illustris is compatible with the allowed range determined by
  \citet{Werk2014} (see their Fig. 11). The high baryon fraction in
  this mass range might also be connected to the finding that the
  galaxy stellar mass function in Illustris is too high for low mass
  galaxies \citep{Genel2014,Vogelsberger2014a}. We also recall that
  \textsc{arepo} leads to more cooling than SPH-based codes for the
  same physics implementation \citep{Vogelsberger2011}.

In \autoref{fig:mass_halos_evo}, we show the redshift evolution of the
fraction of total mass, dark matter and baryon mass bound inside
haloes. We see that the total mass inside haloes is monotonically
increasing with time, reaching \mbox{45.5 \%} at redshift
$z=0$. Between redshift $z=2$ and 1, the mass gain for the
baryonic halo component is less pronounced than for dark matter, and
from redshift $z=1$ to 0 the fraction of baryons inside haloes
even decreases. The onset of this decrease coincides with the
  radio mode AGN feedback becoming important \citep[see][for the
  evolution of the black hole accretion rate in the Illustris
  simulation]{Sijacki2015}. The figure also shows that the dominant
contribution to the baryonic halo budget is from gas at all
redshift. While the fractional contribution of stars to the total
baryon budget increases with time, it reaches only a final value of
\mbox{6.6 \%} at redshift $z=0$.

\subsection{Mass distribution in different dark matter density environments}\label{sec:dm_density}
\begin{figure*}
\centering
\begin{subfigure}{0.45\textwidth}
\includegraphics[width=\textwidth]{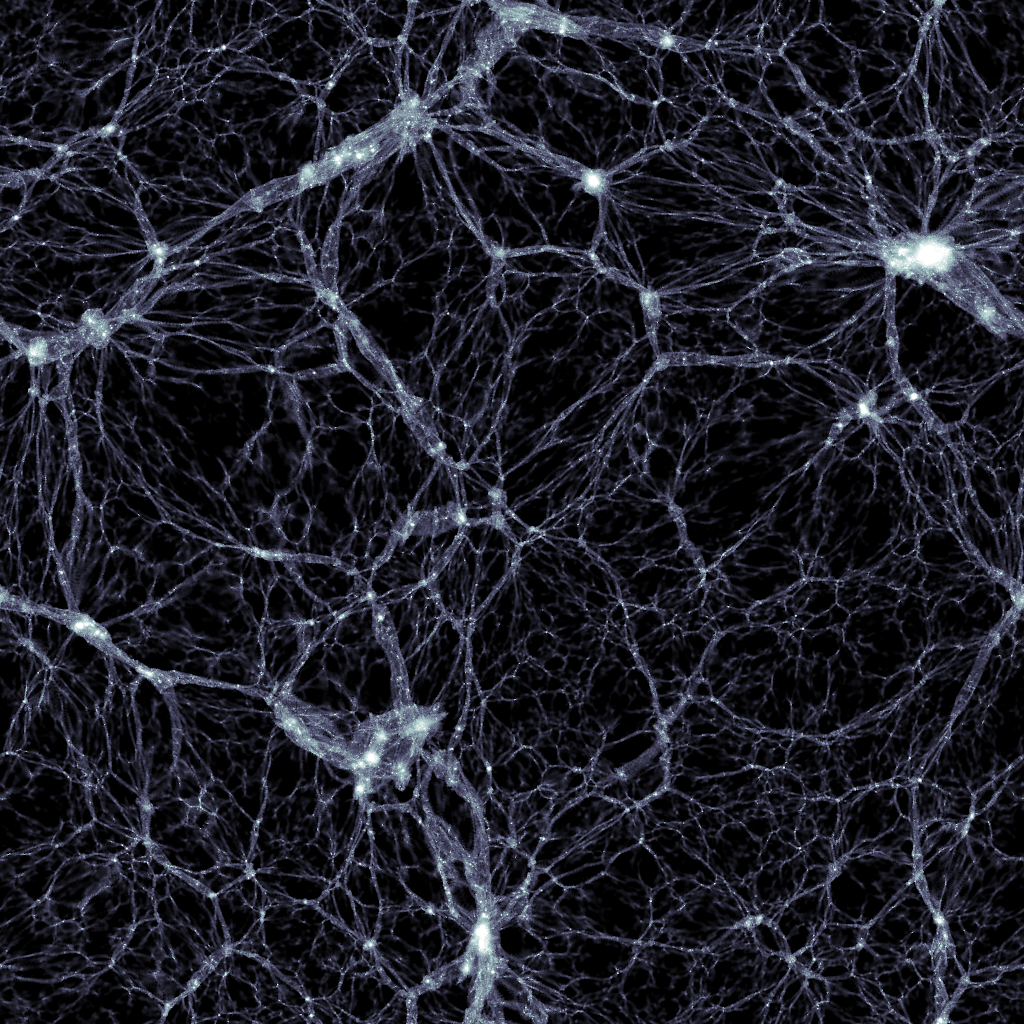}
\caption{dark matter}\label{fig:dm_slice_z0}
\end{subfigure}\qquad\quad\quad
\begin{subfigure}{0.45\textwidth}
\includegraphics[width=\textwidth]{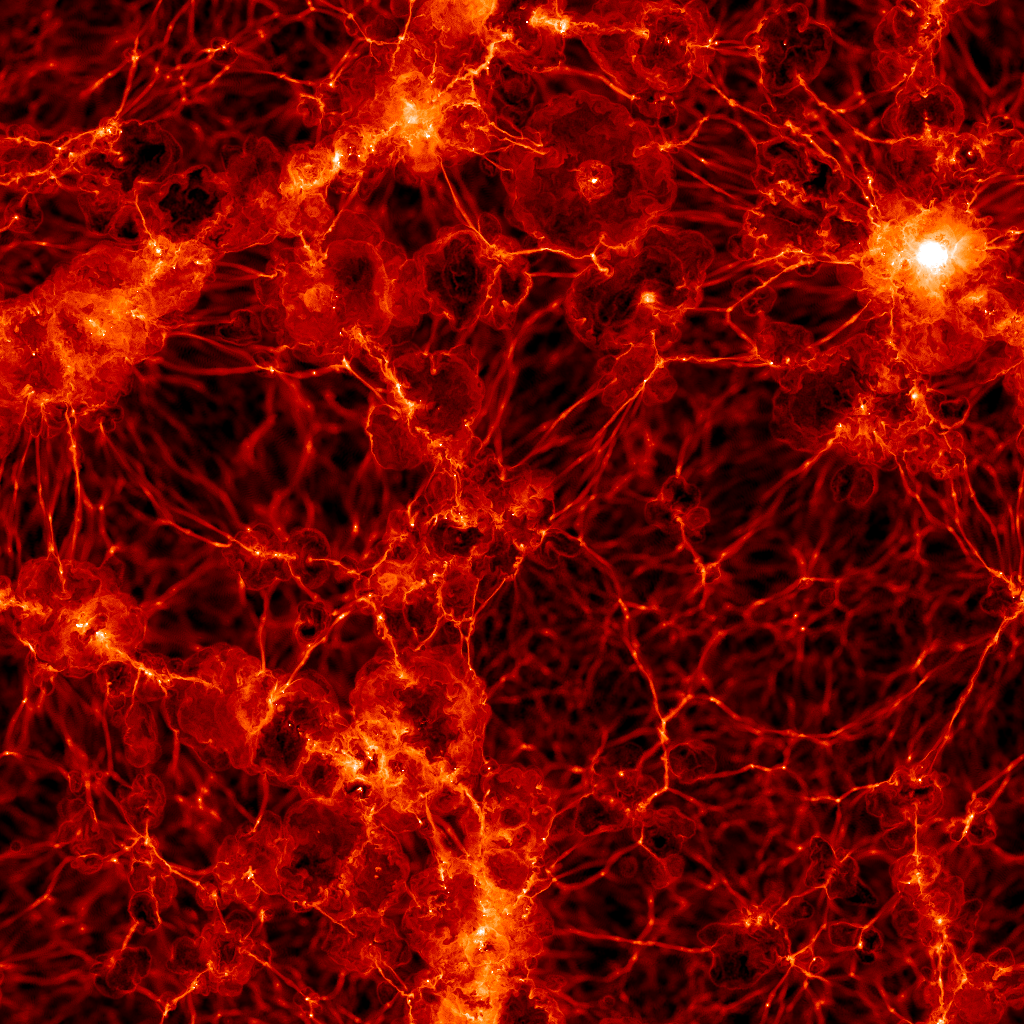}
\caption{baryons}\label{fig:baryon_slice_z0}
\end{subfigure}
\caption{Dark matter and baryon density in a thin slice at $z = 0$. The slice covers the whole (106.5 
Mpc)$^2$ extent of the simulation and has a thickness of 104 kpc (1 cell).}
\label{fig:slice_z0}
\end{figure*}

The formation of large-scale structure is dominated by dark matter. Therefore, in this section we analyse the 
mass distribution with respect to different dark matter density environments. We compute the dark matter 
density through mapping the Illustris data on to a (1024)$^3$ grid covering the whole simulation volume. In 
\autoref{fig:slice_z0} (a) we show a plot of the dark matter density and in \autoref{fig:slice_z0} (b) of the 
baryon density in a slice through the simulation at redshift zero. The slice has an extent corresponding to 
the full 106.5 $\times$ 106.5 Mpc, and a thickness of one cell (104 kpc). For both the dark matter and the 
baryon slice we use a logarithmic colourmap with a range of 4.5 dex. The colourmap boundaries for the baryons 
have been set an order of magnitude lower than for the dark matter.

\begin{figure}
\centering
\includegraphics[width=0.45\textwidth]{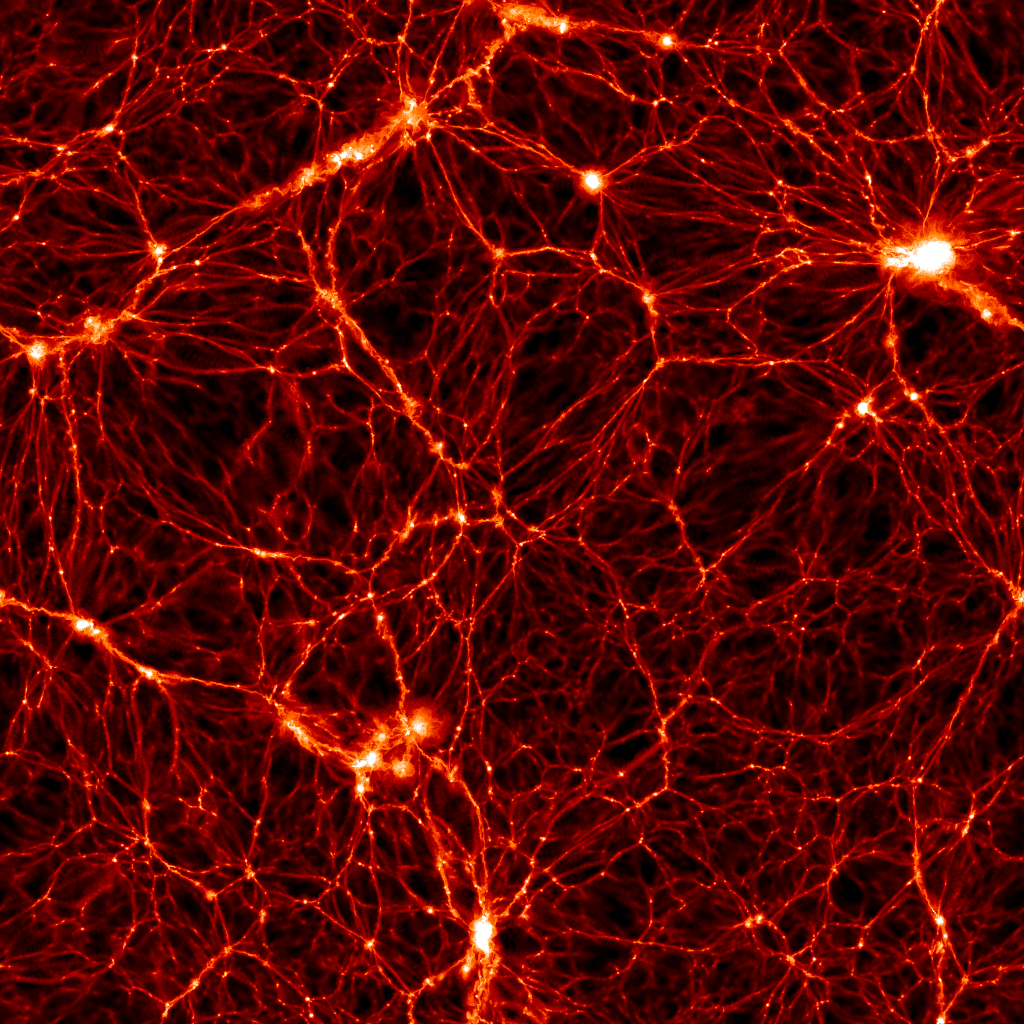}
\caption{Baryon density of the non-radiative simulation (no star formation, feedback or cooling) at z = 0. 
The same slice as in \autoref{fig:slice_z0} is displayed.}
\label{fig:slice_z0NR}
\end{figure}

We note that the baryons appear more extended than the dark matter distribution, especially around high dark 
matter density regions. It is interesting to compare this to \autoref{fig:slice_z0NR}, where we show the 
corresponding baryon density of the non-radiative run. Clearly, in the absence of feedback, the baryons are 
tracing the dark matter well on large scales.


\begin{figure*}
\centering
\begin{subfigure}{0.32\textwidth}
\includegraphics[width=\textwidth]{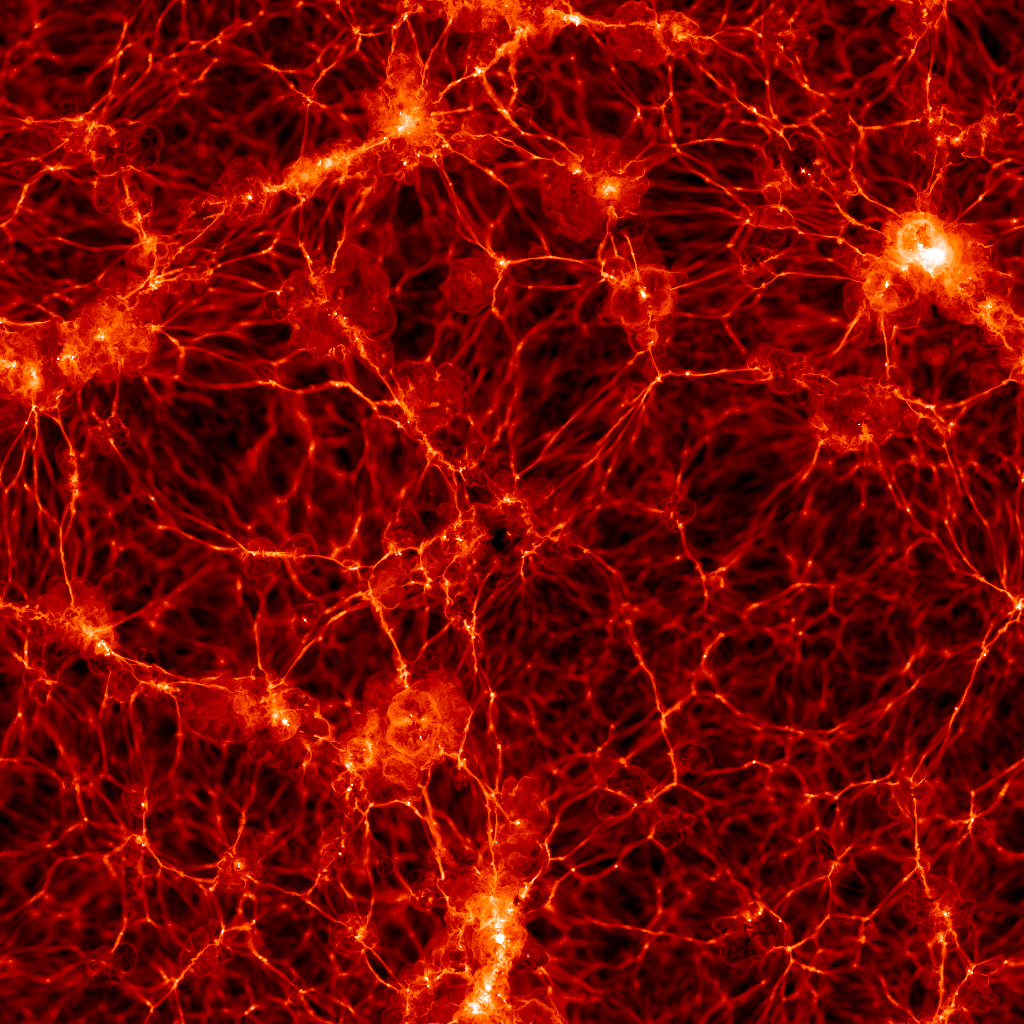}
\caption{z = 0.5}\label{fig:z05_baryons}
\end{subfigure}
\begin{subfigure}{0.32\textwidth}
\includegraphics[width=\textwidth]{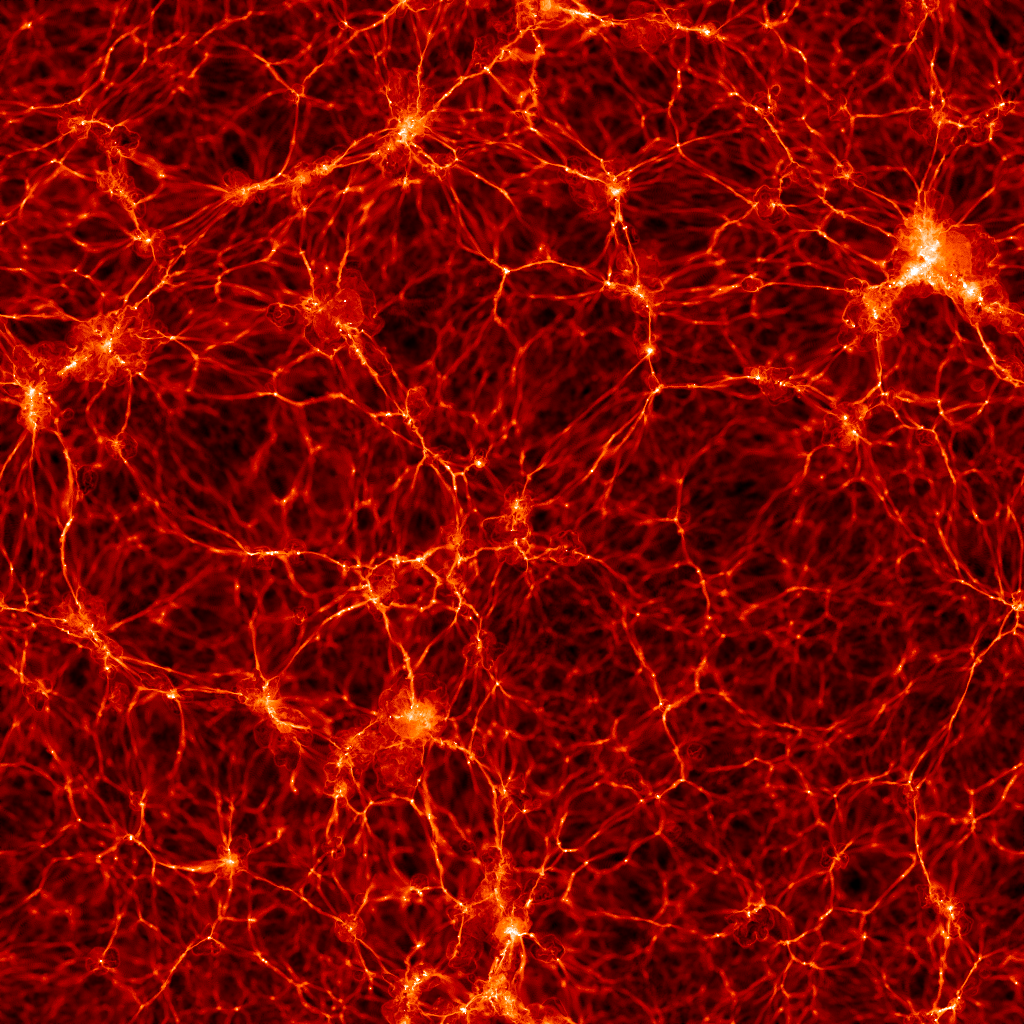}
\caption{z = 1}\label{fig:z1_baryons}
\end{subfigure}
\begin{subfigure}{0.32\textwidth}
\includegraphics[width=\textwidth]{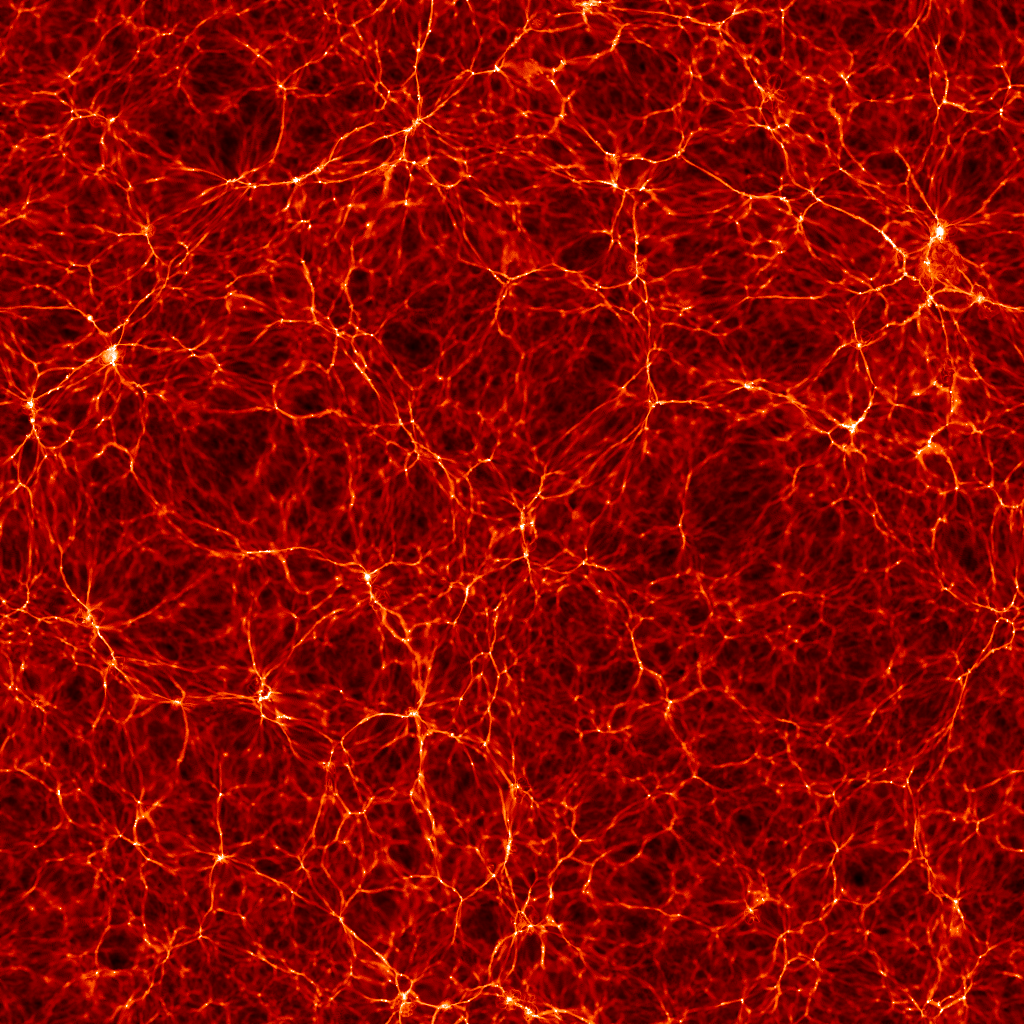}
\caption{z = 2}\label{fig:z2_baryons}
\end{subfigure}
\caption{Redshift evolution of the baryon density. The same slice as in \autoref{fig:slice_z0} has been used.}
\label{fig:slice_redshift_evolution}
\end{figure*}

Comparing the baryon distributions at different redshifts (Figs. \ref{fig:slice_redshift_evolution} and 
\ref{fig:slice_z0}), one can see that the extended baryon distributions around high dark matter density areas 
are less prominent at $z=0.5$ and virtually absent at $z=2$. The diameter of these regions is also increasing 
from $z=0.5$ to 0. We conclude that the extended gas shells originate in gas which has been expelled from 
their host haloes due to radio-mode feedback, which becomes active at low redshift. A rough estimate from the 
diameter of these shells at different redshifts suggests that their expansion speed is between 500 and 1000 
comoving km\,s$^{-1}$.

\subsubsection{Mass distribution according to dark matter density}\label{sec:cosmic_web}
\begin{figure}
\centering
\includegraphics[width=0.49\textwidth]{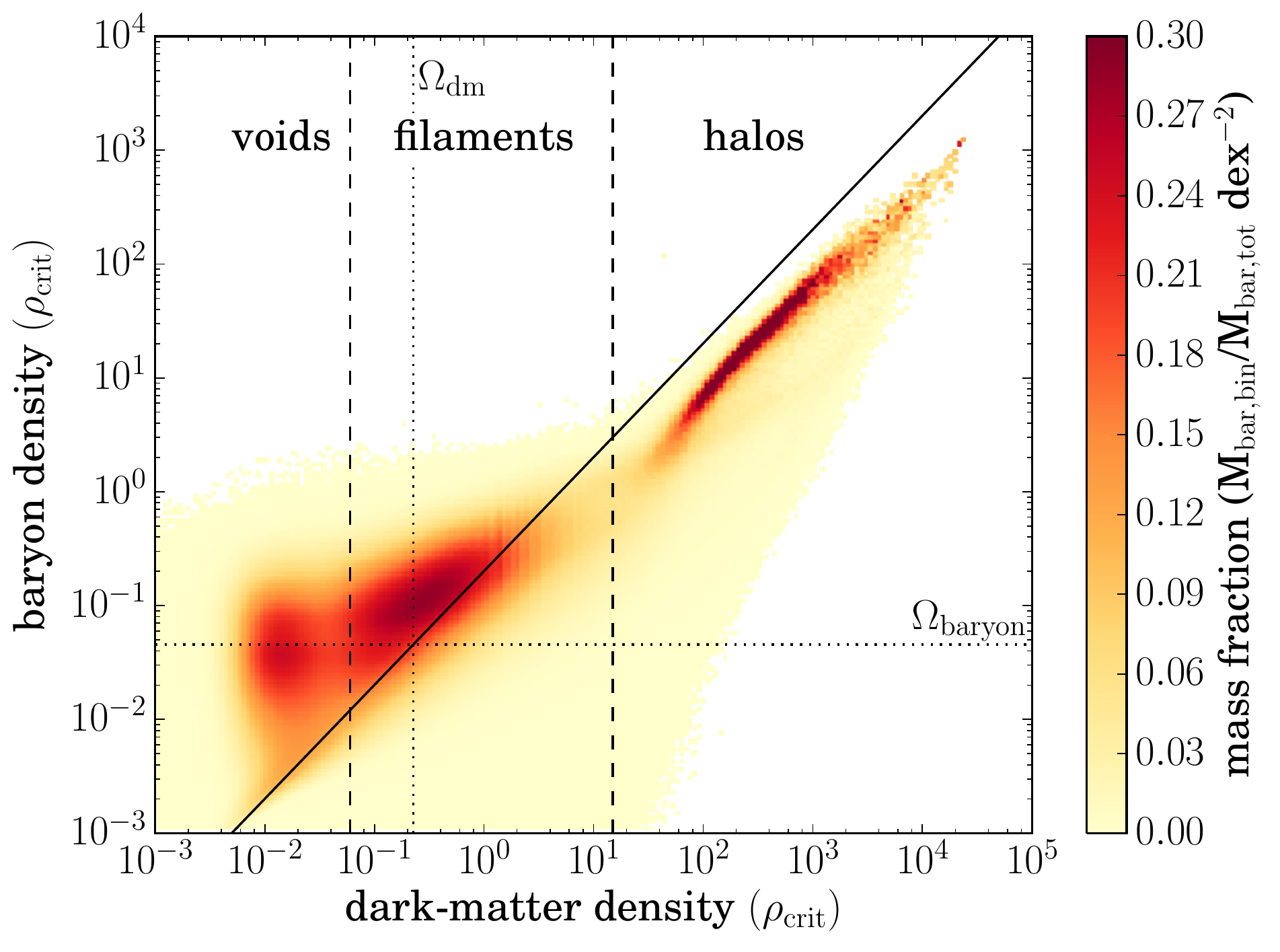}
\caption{Histogram of the contribution of different dark matter and baryon density regions to the total 
baryonic mass. The dashed vertical lines indicate the boundaries of the dark matter density regions we 
defined in \autoref{sec:cosmic_web} to measure the mass in filaments, voids and haloes. The diagonal line 
indicates where a primordial mix of baryons and dark matter would lie.}
\label{fig:baryon_vs_dm_2d}
\end{figure}

In \autoref{fig:baryon_vs_dm_2d}, we present the fraction of the baryons at a given baryon and dark matter 
density. We binned the data in baryon and dark matter density using 200 $\times$ 200 bins. The diagonal line 
gives \OmegaB/\OmegaDM$\times\rho_\textrm{\small dm}$, which is where the baryons would lie if they traced 
dark matter perfectly.
The colourmap shows the contribution of a bin to the total baryon mass. We find that there is a significant 
fraction of mass at high dark matter and high baryon densities, which corresponds to gas in haloes. A large 
fraction of baryons is at intermediate dark matter densities around the primordial dark matter density 
\OmegaDM\rhocrit. However, there are also many baryons at the lowest dark matter densities, the region we 
labelled `voids'.
\begin{figure*}
\centering
\begin{subfigure}{0.45\linewidth}
\includegraphics[width=\textwidth]{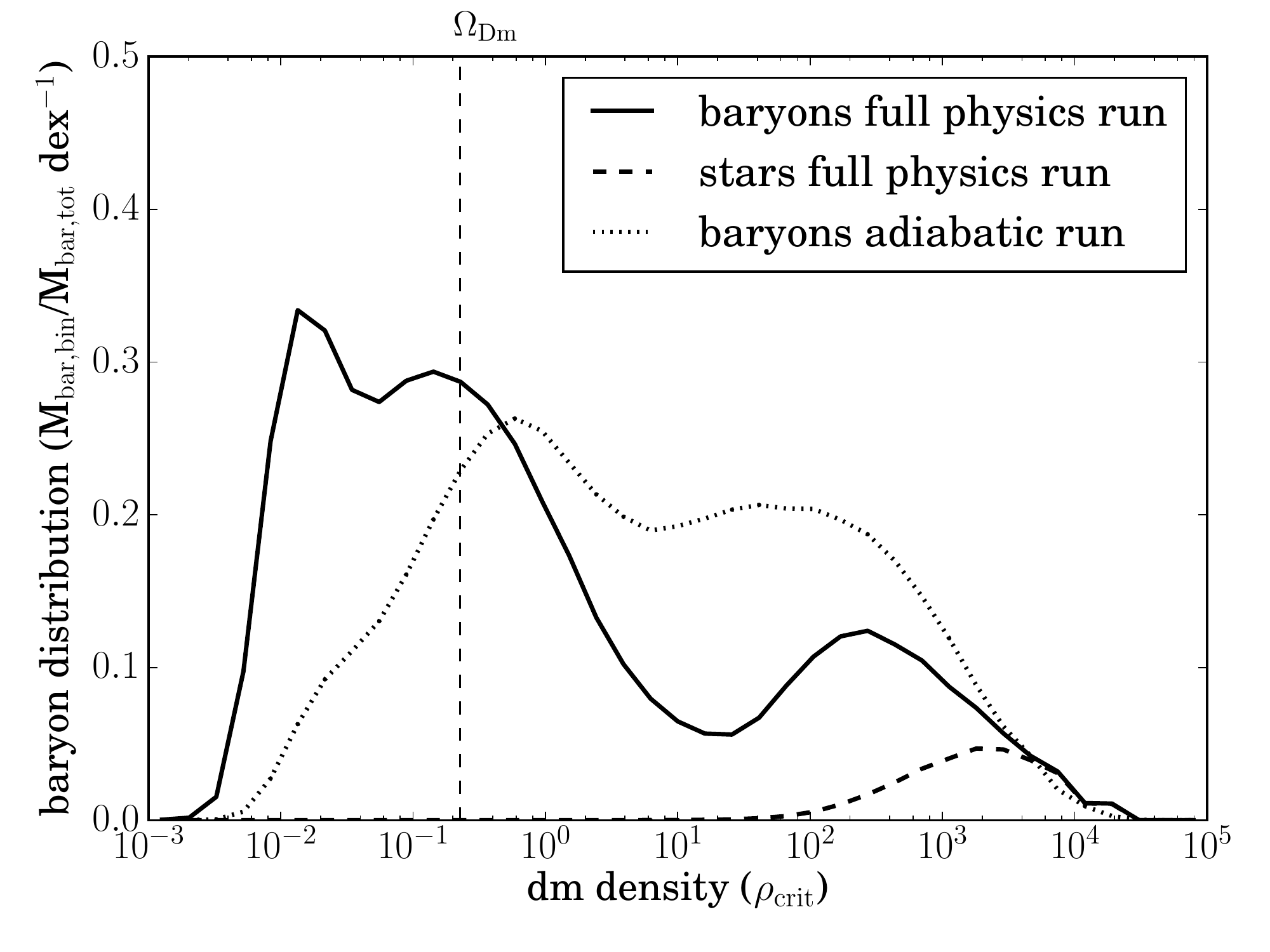}
\caption{}\label{fig:mass_distro_baryons_z0}
\end{subfigure}
\begin{subfigure}{0.45\linewidth}
\includegraphics[width=\textwidth]{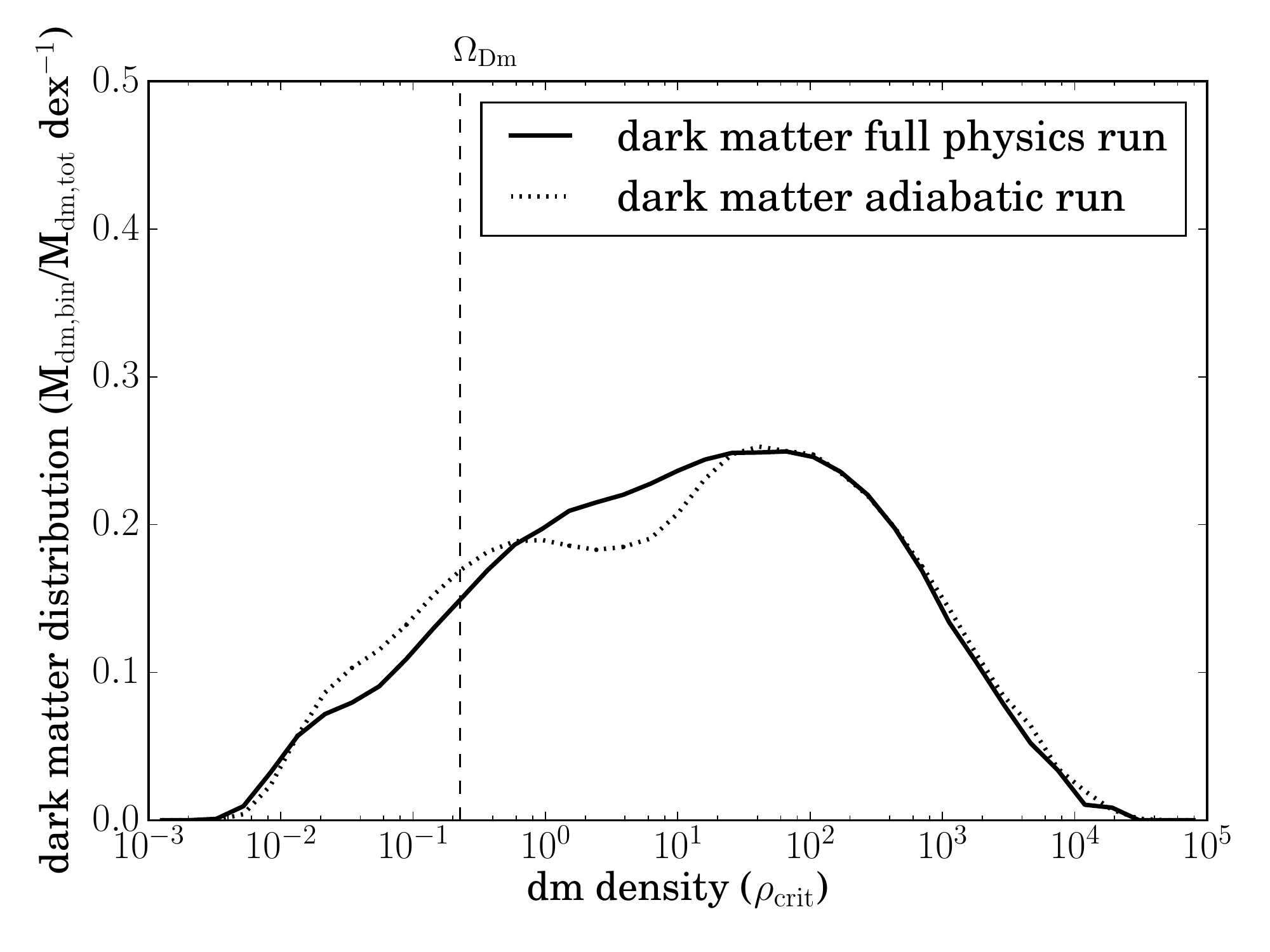}
\caption{}\label{fig:mass_distro_dm_z0}
\end{subfigure}
\caption{Baryon and dark matter mass distribution with respect to the dark matter density at $z=0$. In 
(\subref{fig:mass_distro_baryons_z0}), it is shown how much baryonic mass resides in an environment with a 
given dark matter density. The $y$-axis shows the fraction of baryonic mass in a dark matter density bin to 
the total baryonic mass in the simulation volume. The solid line represents the data from the full physics 
run and the dotted line shows the values for the non-radiative run. The dashed line gives the contribution of 
stars to the baryons in the full physics run. In (\subref{fig:mass_distro_dm_z0}), we show the analogous plot 
for the dark matter mass.}
\label{fig:mass_distro_z0}
\end{figure*}

It is interesting to observe the differences between the full physics and the non-radiative runs in 
\autoref{fig:mass_distro_z0}, which shows the distribution of mass according to the dark matter density. We 
binned the data in dark matter density and then measured the mass contribution of a bin to the total baryonic 
or dark matter mass in the simulation volume. We see that the dark matter distribution in (b) is very similar 
for the full physics and the non-radiative run. However, there are notable differences for the baryons. For 
the full physics simulation, there is less mass at higher dark matter densities and significantly more mass at 
the lowest dark matter densities. This again is due to the feedback processes present only in the full 
physics simulation, which expel matter from the haloes out into dark matter voids.

\subsubsection{Dark matter density and the cosmic web}
\begin{figure*}
\centering
\begin{subfigure}{0.32\textwidth}
\includegraphics[width=\textwidth]{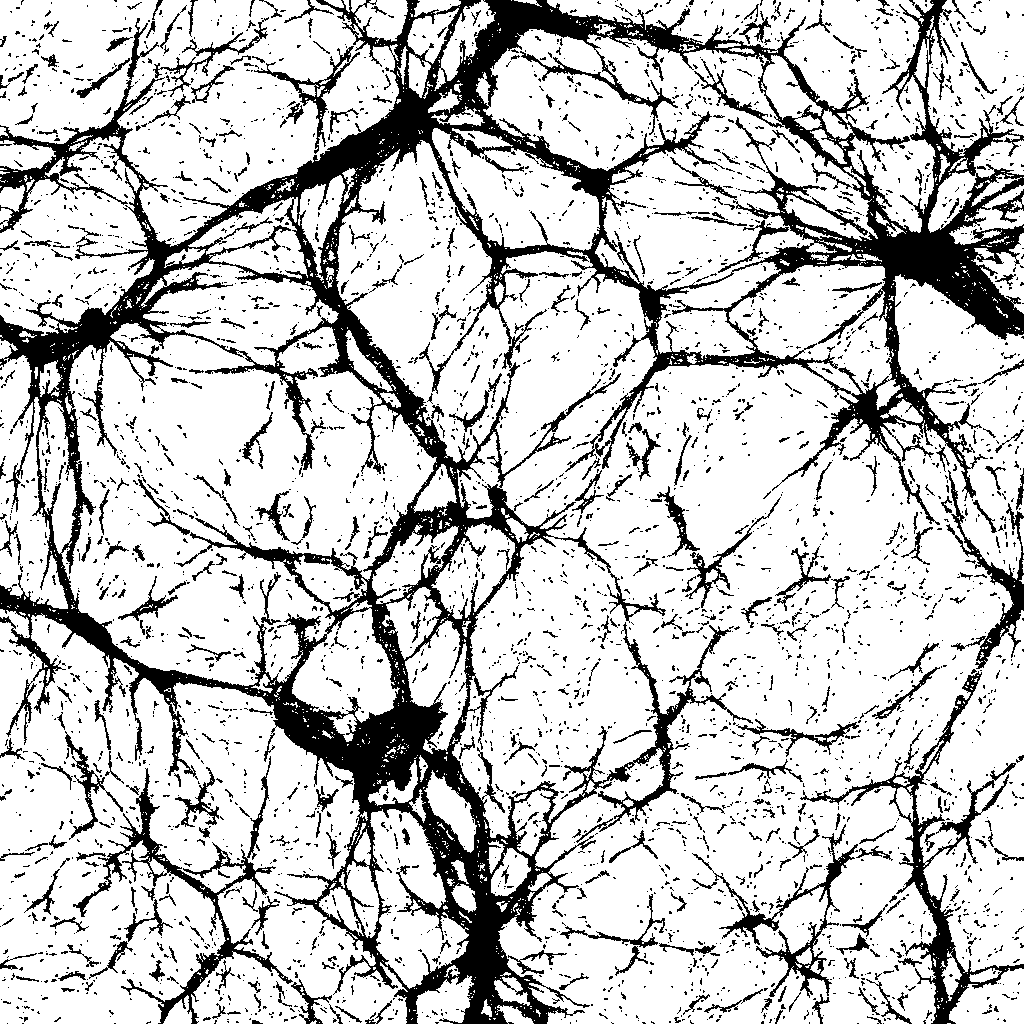}
\caption{voids}
\end{subfigure}
\begin{subfigure}{0.32\textwidth}
\includegraphics[width=\textwidth]{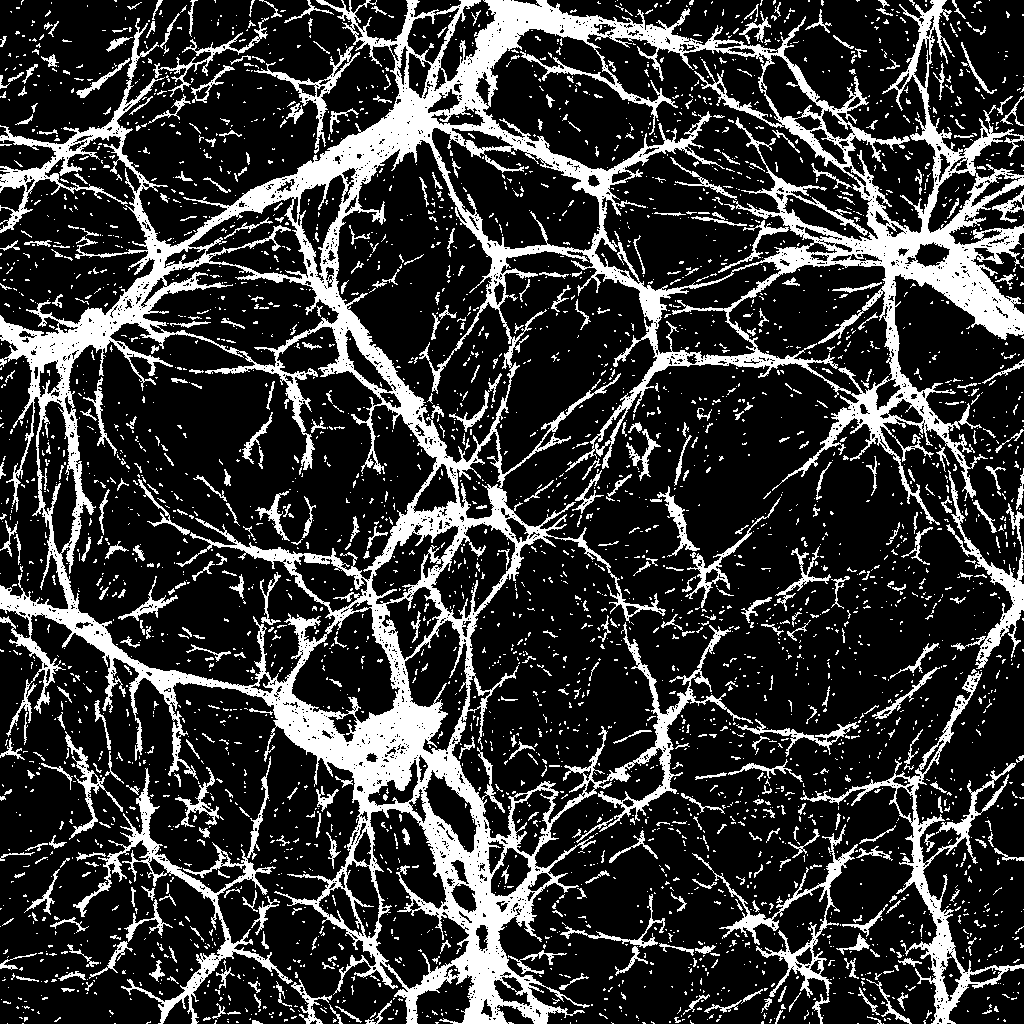}
\caption{filaments}
\end{subfigure}
\begin{subfigure}{0.32\textwidth}
\includegraphics[width=\textwidth]{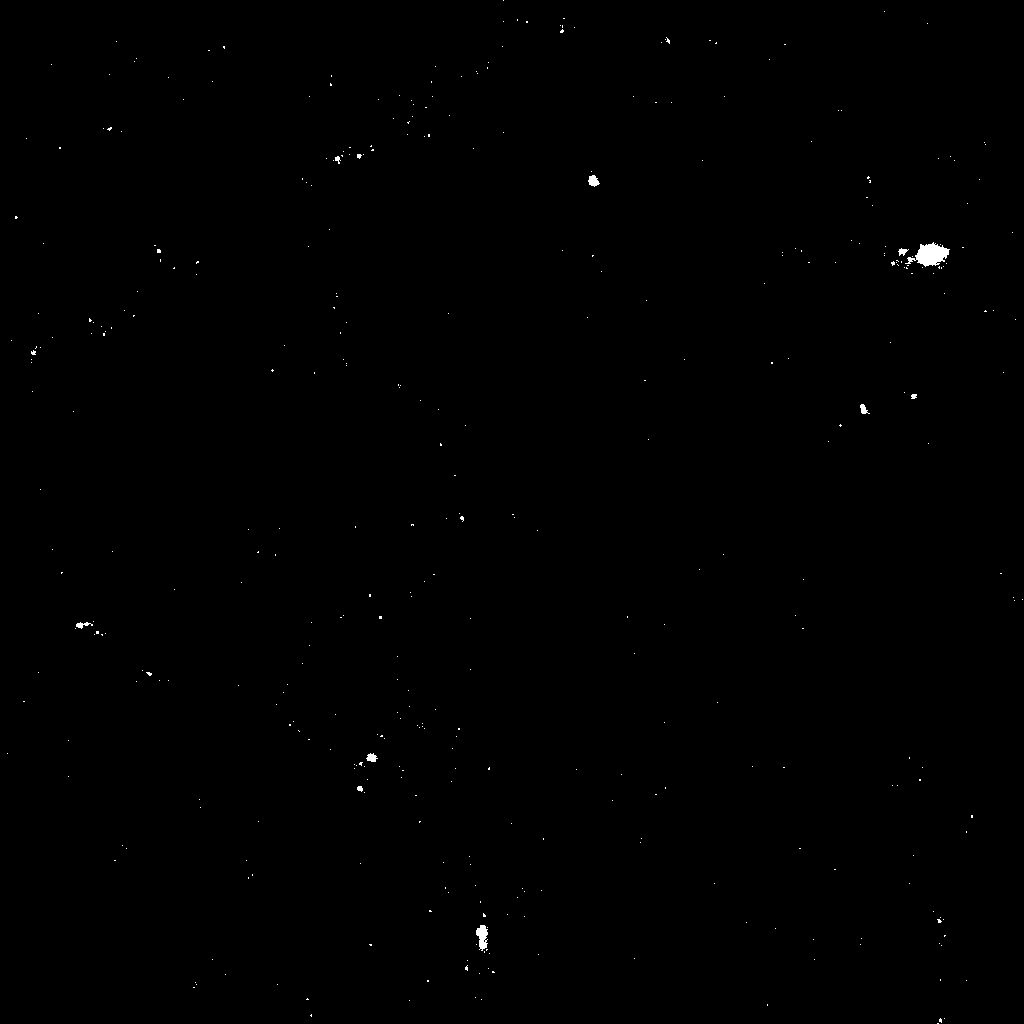}
\caption{haloes}
\end{subfigure}
\caption{Spatial regions corresponding to the dark matter density ranges defined in \autoref{table:regions}. 
If a cell's dark matter density is inside the corresponding region it is drawn in white, otherwise it is 
black. The same slice as in \autoref{fig:slice_z0} has been used.}
\label{fig:dm_density_partition}
\end{figure*}

The primary constituents of the large-scale structure differ in their dark matter density. Therefore we can 
use the dark matter density itself as a rough proxy to split up the simulation volume into regions 
corresponding to haloes, filaments and voids. We thus define the categories haloes, 
filaments and voids by assigning to them a certain dark matter density range\footnote{The 
dark matter densities we use in this analysis are average densities over a volume of (104 kpc)$^3$.}. By 
summing the grid cells which fall into the respective dark matter density ranges we can measure the mass and 
volume of those regions. We show the resulting mass and volume fraction and the density range used for this 
classification in \autoref{table:regions}.
The spatial distribution of the categories in \autoref{table:regions} is shown in Fig. 
\ref{fig:dm_density_partition}. We show the same slice as in \autoref{fig:slice_z0}, and use a white colour 
for those cells that belong to the corresponding category.

\begin{table*}
\caption{Dark matter mass, baryonic mass and volume fraction in haloes, filaments and voids at z=0. The 
categories have been defined through dark matter density ranges. We also added a category `ejected material' 
which corresponds to baryons inside the `voids' region which have a temperature T$> 6\times 10^4$\,K. The 
spatial regions to which these dark matter density regions correspond to are shown in 
\autoref{fig:dm_density_partition}.}
\label{table:regions}
\centering
\begin{tabular}{cccccc}
\hline
&dark matter density& \% of total & \% of total & \% of total & \% of total\\  
 component&region (\rhocrit)  &  dark matter mass & baryonic mass & mass & volume \\ \hline
haloes & $ >$ 15 & 49.2 \% & 23.2 \% & 44.9 \% & 0.16 \% \\
filaments & 0.06 - 15 & 44.5 \% & 46.4 \% & 44.8 \%& 21.6 \% \\
voids & 0 - 0.06 & 6.4 \% & 30.4 \% & 10.4 \% & 78.2 \% \\
ejected material& \multirow{2}{*}{0 - 0.06} & \multirow{2}{*}{2.6 \%}  & \multirow{2}{*}{ 23.6 
\%}&\multirow{2}{*}{6.1 \%} &\multirow{2}{*}{30.4 \%}\\
 inside voids & & & &\\ \hline
\end{tabular}
\end{table*}

The `haloes' are defined to have a dark matter density higher than 15\rhocrit. This density threshold has been
chosen so that the halo region contains approximately the same amount of dark matter as the haloes of the halo finder.
Thus, by construction, we find \mbox{49 \%} of the dark matter in haloes. But also the amount of baryons, which corresponds to
\mbox{23 \%}, is in good agreement with \autoref{table:mass_in_halos}. The volume fraction of the haloes is \mbox{0.16 \%}. The chosen
threshold is also consistent with the resolution of the grid on which we calculate the density and the mass resolution of 
Illustris: a halo of 2 $\times 10^9$ \Msol\ would result in an overdensity of 15$\times$\rhocrit , if all of its mass 
falls into a single grid cell. According to the \textsc{subfind} halo catalogue, haloes smaller 
than 2$\times 10^9$ \Msol\ still host \mbox{4.8 \%} of the dark matter. However, many of those haloes will 
be subhaloes and thus contribute their mass to the halo category. If we define the halo category to have densities higher than 30\rhocrit, we find that it 
hosts \mbox{41.7 \%} of the dark matter and \mbox{21.6 \%} of the baryons, and all 
of the stars. This density cut at 30\rhocrit\ corresponds to the mass of a 4.2 $\times 10^9$ \Msol\ halo in 
one cell. If we compare this value with the second row in \autoref{table:mass_in_halos} 
\mbox{($M_\textrm{tot} >5\times 10^9$ \Msol)} we again find that the value obtained using the dark matter 
density method is consistent with the value of the \textsc{subfind} halo catalogue. We want to emphasize that
the threshold density for the haloes is depending on the resolution of the grid, as it is the average density
in one grid cell, and not the physical density at the outskirts of haloes.

The `filaments' region, which we defined to have dark matter densities between 0.06 to 15 \rhocrit, hosts 
\mbox{44.5 \%} of the dark matter and \mbox{46.4 \%} of the baryons. Its volume makes up \mbox{21.6 \%} of 
the total simulation volume.
The filaments span a large density range in our definition. At the higher density end of this definition, 
between 5 and 15 \rhocrit, the mass is mainly in ring-like regions around the bigger haloes, and no 
filamentary structure is visible in this density regime. This circumhalo region hosts about \mbox{11.1 \%} 
of the dark matter and \mbox{3.3 \%} of the baryons. The density boundary to the voids is
motivated by \autoref{fig:baryon_vs_dm_2d}, as the large bulk of mass in the denser parts of the filaments 
extends down to this value. Also, inspecting \autoref{fig:dm_density_partition}, this density range seems to correspond
to what one would visually label as filaments. However, a different density range could as well be justified.

\begin{figure}
\centering
\includegraphics[width=0.32\textwidth]{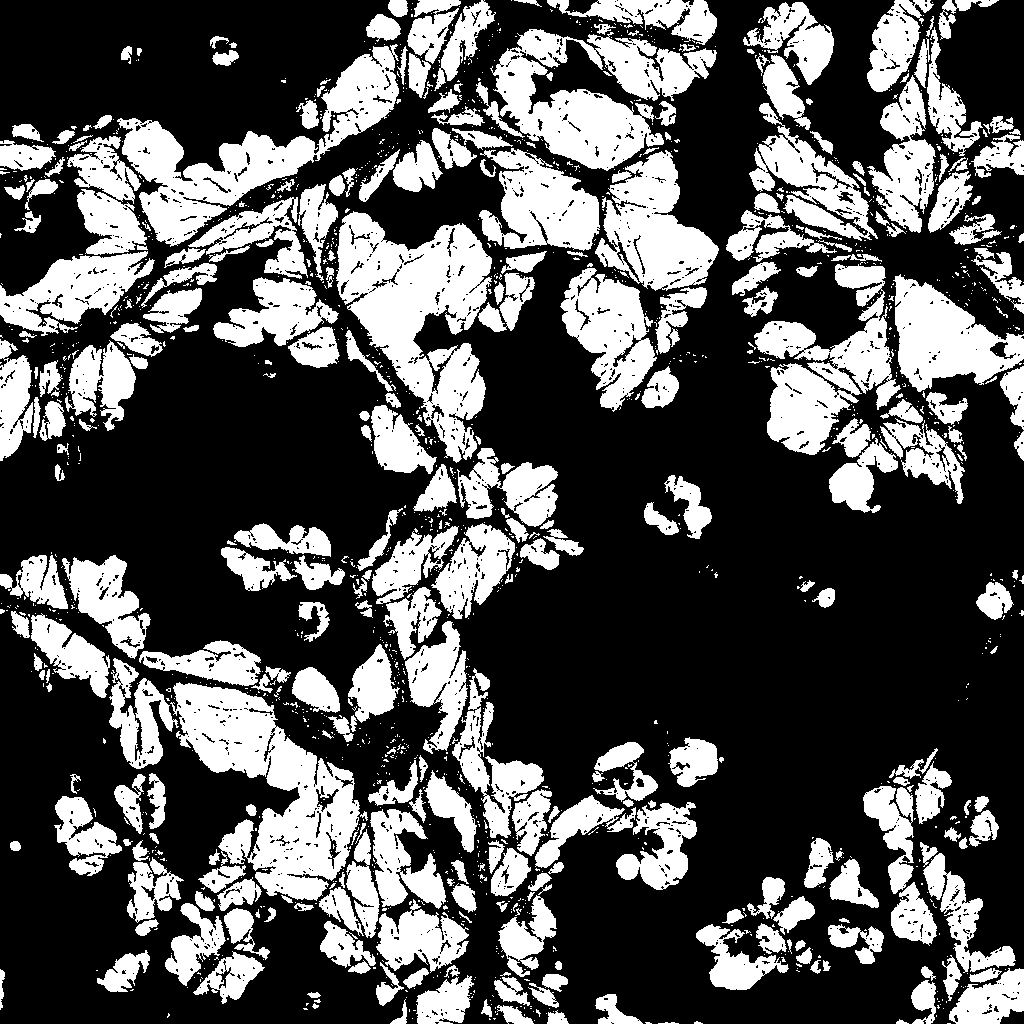}
\caption{Spatial region corresponding to the `ejected baryons' range of \autoref{table:regions}. In addition 
to the dark matter density cut, we here require a temperature higher than \mbox{$T> 6\times10^4 $K}.}
\label{fig:dm_density_partition_ejected}
\end{figure}

\begin{figure}
\centering
\includegraphics[width=0.42\textwidth]{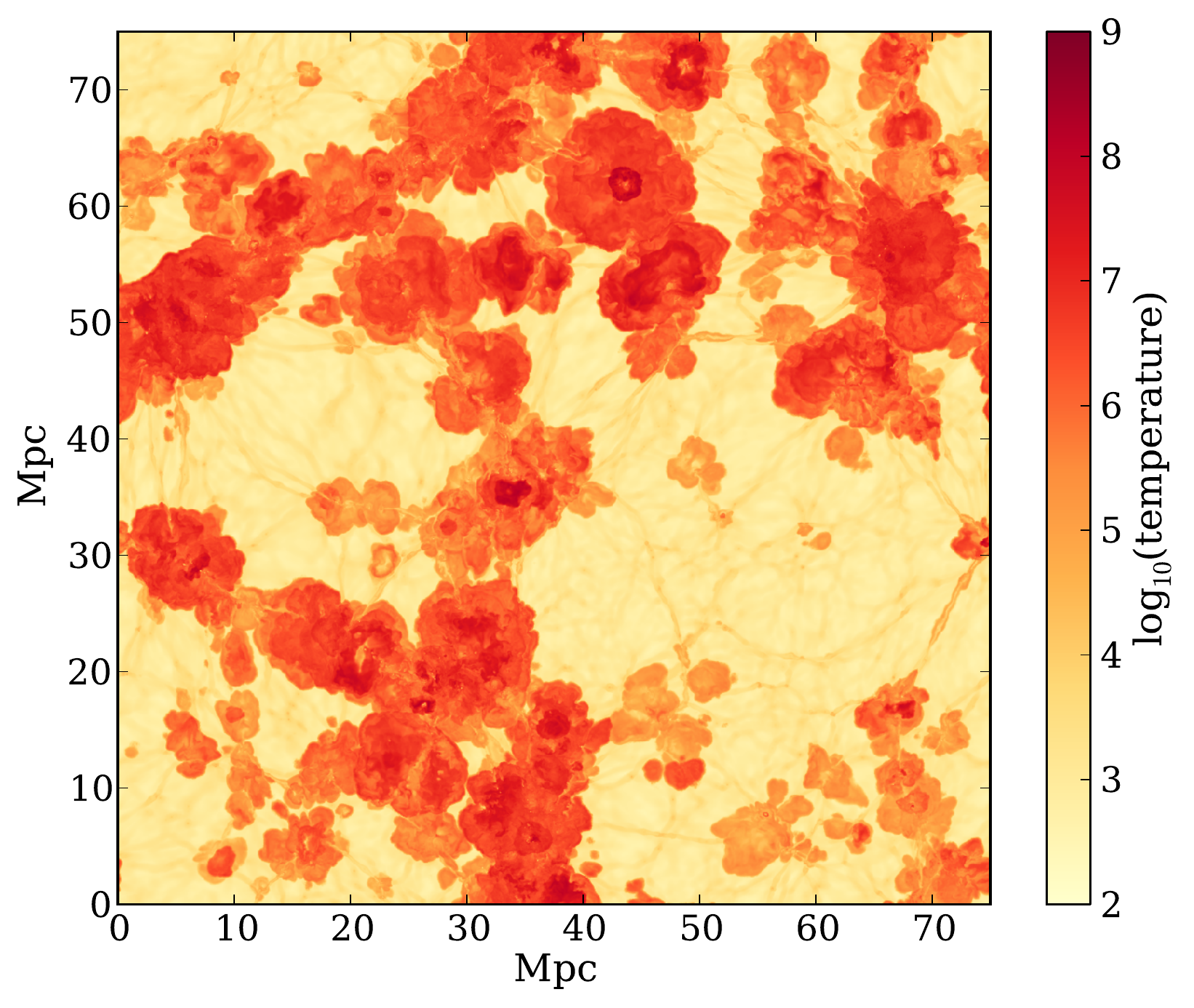}
\caption{Temperature field of the gas in the slice of \autoref{fig:slice_z0}. We see that the gas forms hot 
bubbles around the haloes and filaments.}
\label{fig:temperature_slice}
\end{figure}

\begin{figure}
\centering
\includegraphics[width=0.44\textwidth]{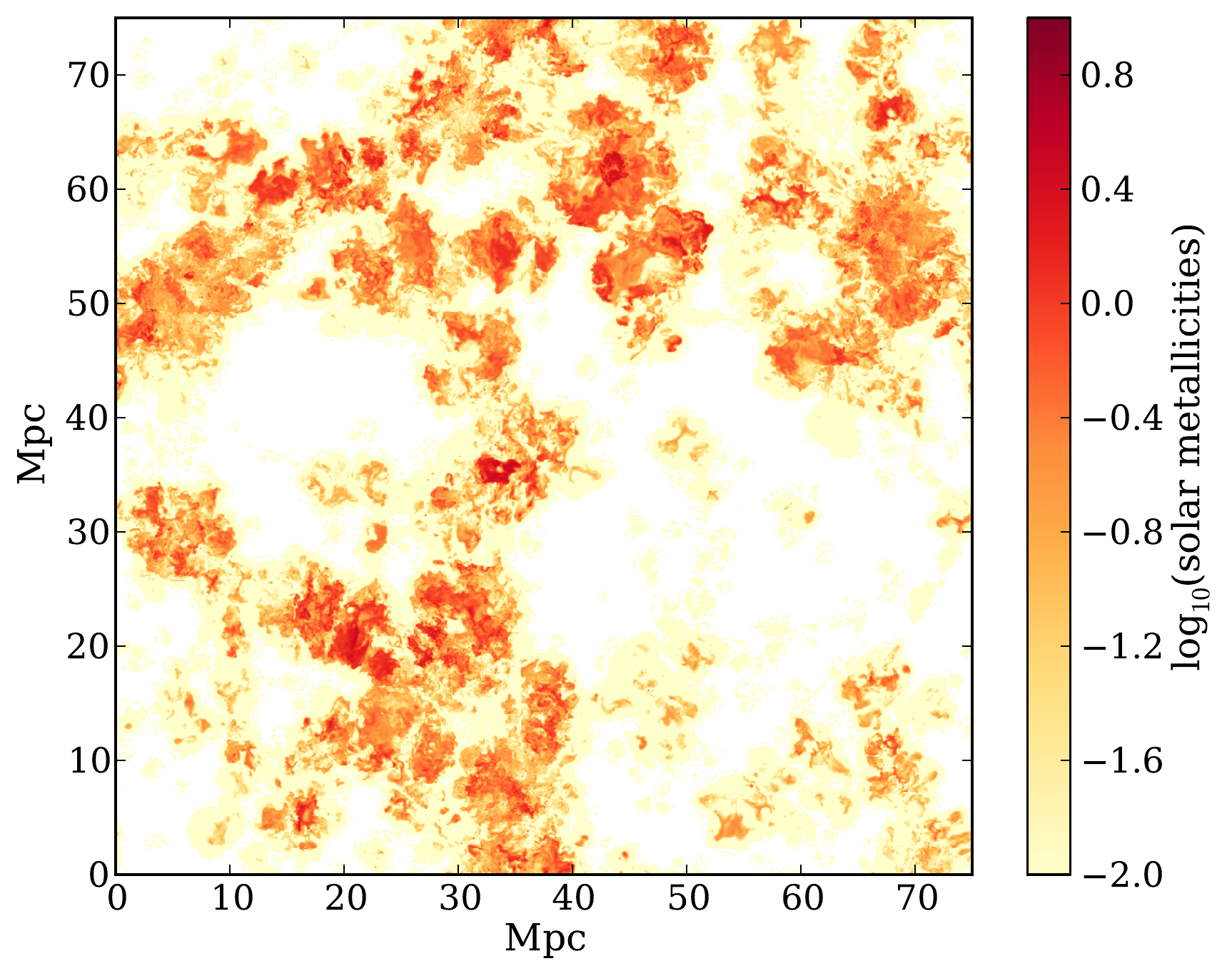}
\caption{Plot of the gas metallicity for the slice which has been used in \autoref{fig:slice_z0}. The 
extended baryon distributions which protrude into the dark matter voids are rich in metals, suggesting that 
feedback processes are the cause of the extended distribution.}
\label{fig:metallicity_slice}
\end{figure}

The `voids' (dark matter densities between 0 and \mbox{0.06 \rhocrit}) contain only \mbox{6.4 \%} of the dark 
matter but \mbox{30.4 \%} of the baryons. The volume of the voids makes up \mbox{78.2 \%} of the total 
simulation volume. In that respect it is worthwhile to examine Figs. \ref{fig:temperature_slice} and 
\ref{fig:metallicity_slice}, where we show the temperature and the metallicity of the gas. If we compare the 
temperature and metallicity maps with \autoref{fig:slice_z0} and \autoref{fig:dm_density_partition}, we see 
that the majority of the baryons in voids is composed of warm to hot gas enriched by metals. Those baryons 
have most likely been ejected from the haloes through feedback. Because the ejected material has a higher 
temperature than the other baryons in voids, we can use the temperature to discriminate between baryons 
naturally residing in voids and baryons which have been transported there. We define an additional `ejected 
material' region in \autoref{table:regions} which is defined as 
having a temperature higher than $6\times 10^4$\,K in addition to the dark matter density cut. With 
\mbox{23.6 \%} of the total baryons, this `ejected material' region is responsible for most of the baryons in 
dark matter voids. In \autoref{fig:dm_density_partition_ejected}, the spatial region corresponding to the 
ejected material is plotted; note that it fills about 40 \% of the voids. We should note though, that the 
ejected mass most likely heats some of the baryons already present in the voids. Therefore, we have probably 
overestimated the ejected mass in voids. However, through following the redshift evolution of the mass in 
voids we can give an estimate of the associate uncertainty, as we discuss below. We note that our findings 
for the volume fractions are generally in good agreement with simulations by \citet{Cautun2014}.

\subsubsection{Redshift evolution of matter and metals in haloes, filaments and voids}

\begin{figure*}
\centering
\begin{subfigure}{0.44\textwidth}
\includegraphics[width=\textwidth]{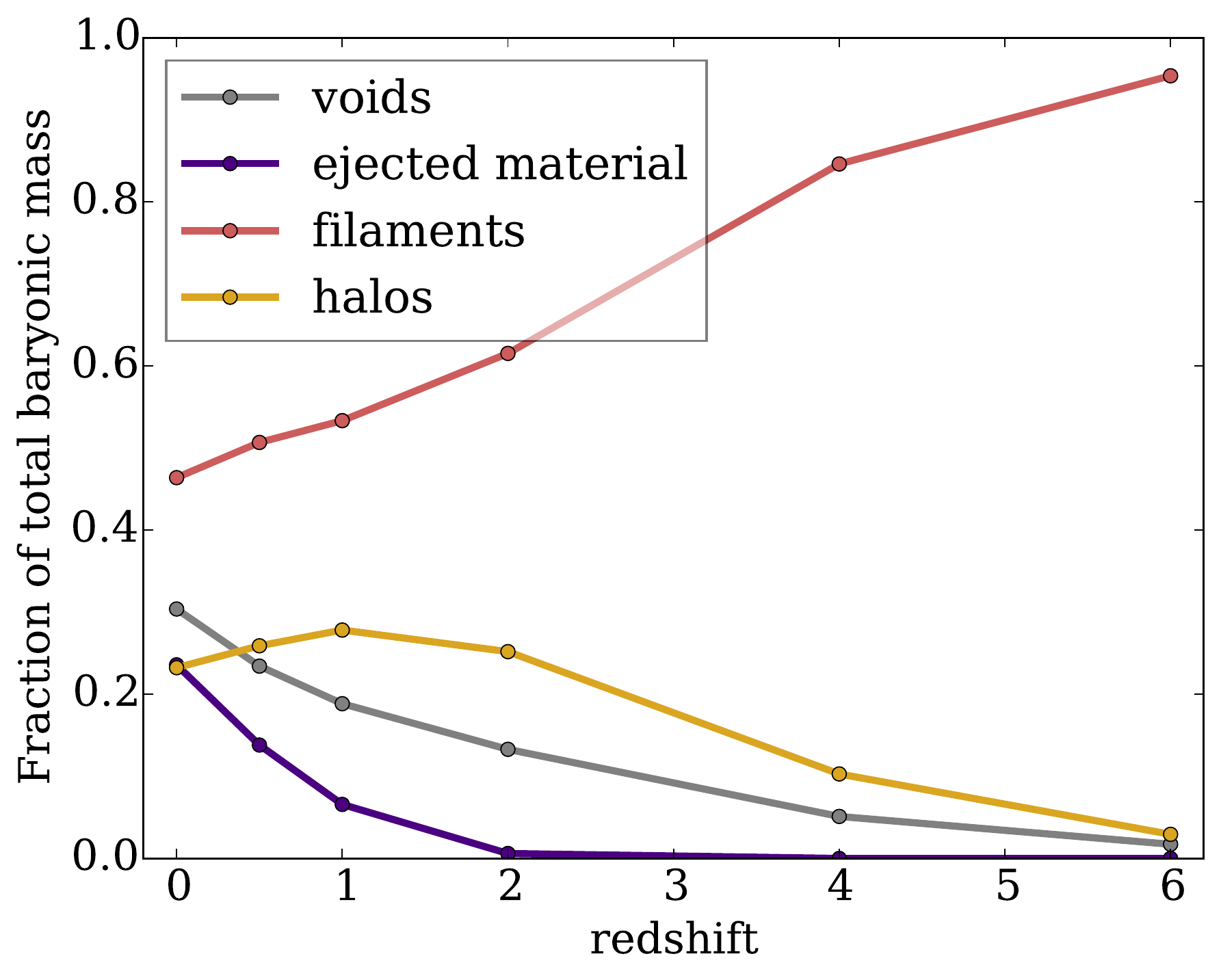}
\caption{}
\end{subfigure}
\begin{subfigure}{0.44\textwidth}
\includegraphics[width=\textwidth]{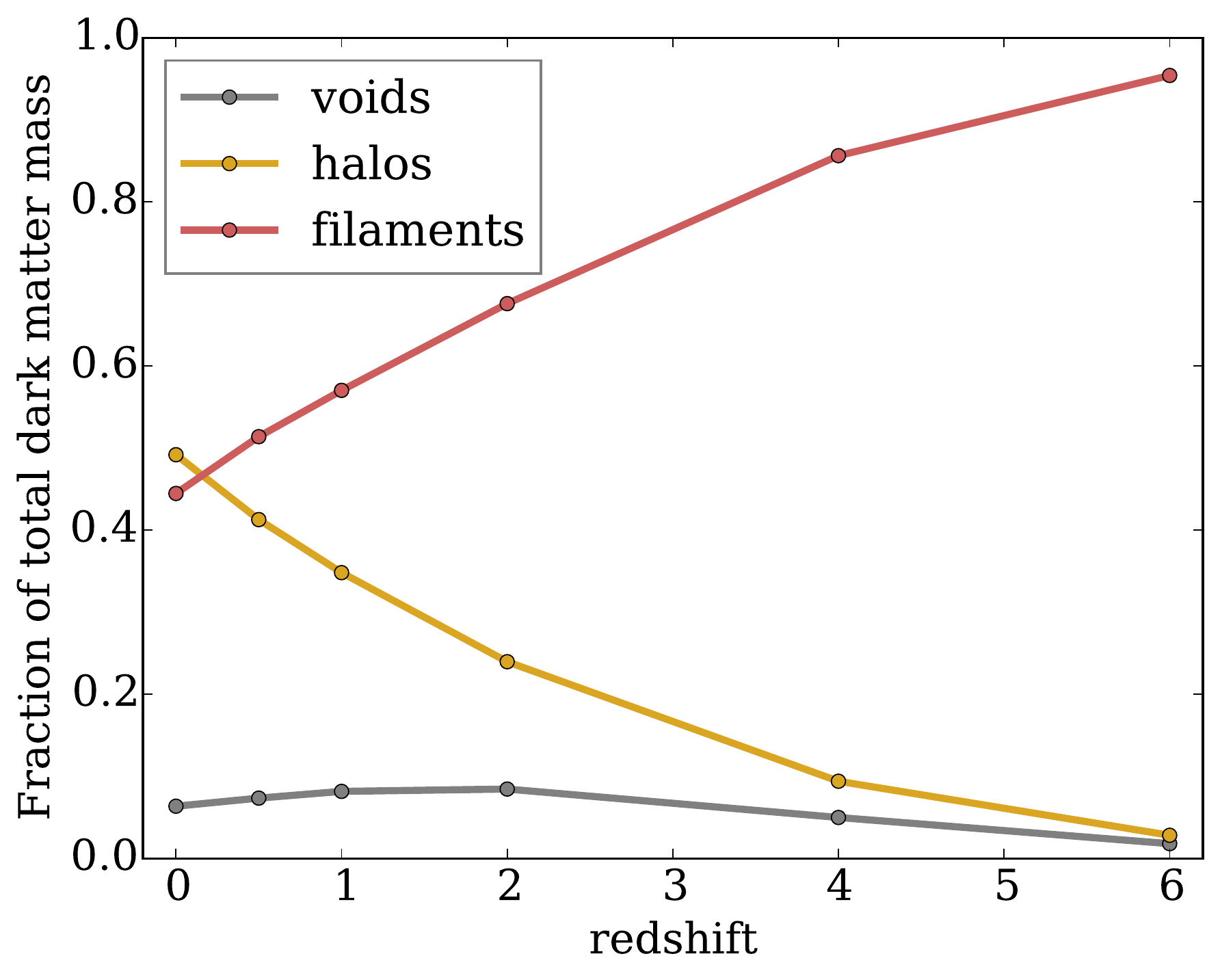}
\caption{}
\end{subfigure}
\caption{Redshift evolution of the contribution of haloes, voids and filaments to the total baryonic mass (a) 
and the total dark matter mass (b). While the evolution of the baryons and the dark matter is similar at 
redshifts higher than two, at low redshifts the haloes, lose baryons due to feedback processes while the dark 
matter in haloes is still growing.}
\label{fig:partition_evo}
\end{figure*}

By applying the same dark matter density cuts at different redshifts, we can study the time evolution of the 
values reported in \autoref{table:regions}. This is done in \autoref{fig:partition_evo}, where we show how 
the baryons and dark matter divide into haloes, filaments and voids as a function of time.
In \autoref{fig:partition_evo} (a) we see that, starting at redshift $z=2$, feedback begins to efficiently 
remove gas from haloes. At first, this only slows down halo growth, but after a redshift of $z=1$, it reduces 
the amount of baryons in haloes. In \autoref{fig:partition_evo} (b) we see that the dark matter haloes, 
unaffected by feedback, continue to grow at the expense of the filaments. At high redshifts, the dark matter 
was distributed homogeneously with a density of \OmegaDM\rhocrit, and thus falls into the `filament' 
category. The underdense regions of the voids were only created as matter from less dense regions was pulled 
into denser regions. Thus the fraction of dark matter in voids is increasing from $z=6$ to 2. After 
$z=2$, the amount of dark matter in voids is slowly decreasing due to accretion on to filaments. The baryons 
show a similar behaviour from $z=6$ to 2. However, starting at a redshift of $z=2$ ejected material is 
also transported into the voids, thus increasing their baryonic 
content. We see in \autoref{fig:partition_evo} (a) that the mass increase of the `ejected material' is higher 
than the mass increase of the `voids'. The most likely explanation is that the ejected mass heats gas already 
present in the voids, which means that we overestimate the mass of the ejected material with our density and 
temperature cut. If we assume that in the absence of feedback the baryons would show the same relative 
decrease from $z=2$ to 0 as the dark matter, we would need to correct the value of the ejected material down 
to \mbox{20 \%}.

\begin{figure}
\centering
\includegraphics[width=0.42\textwidth]{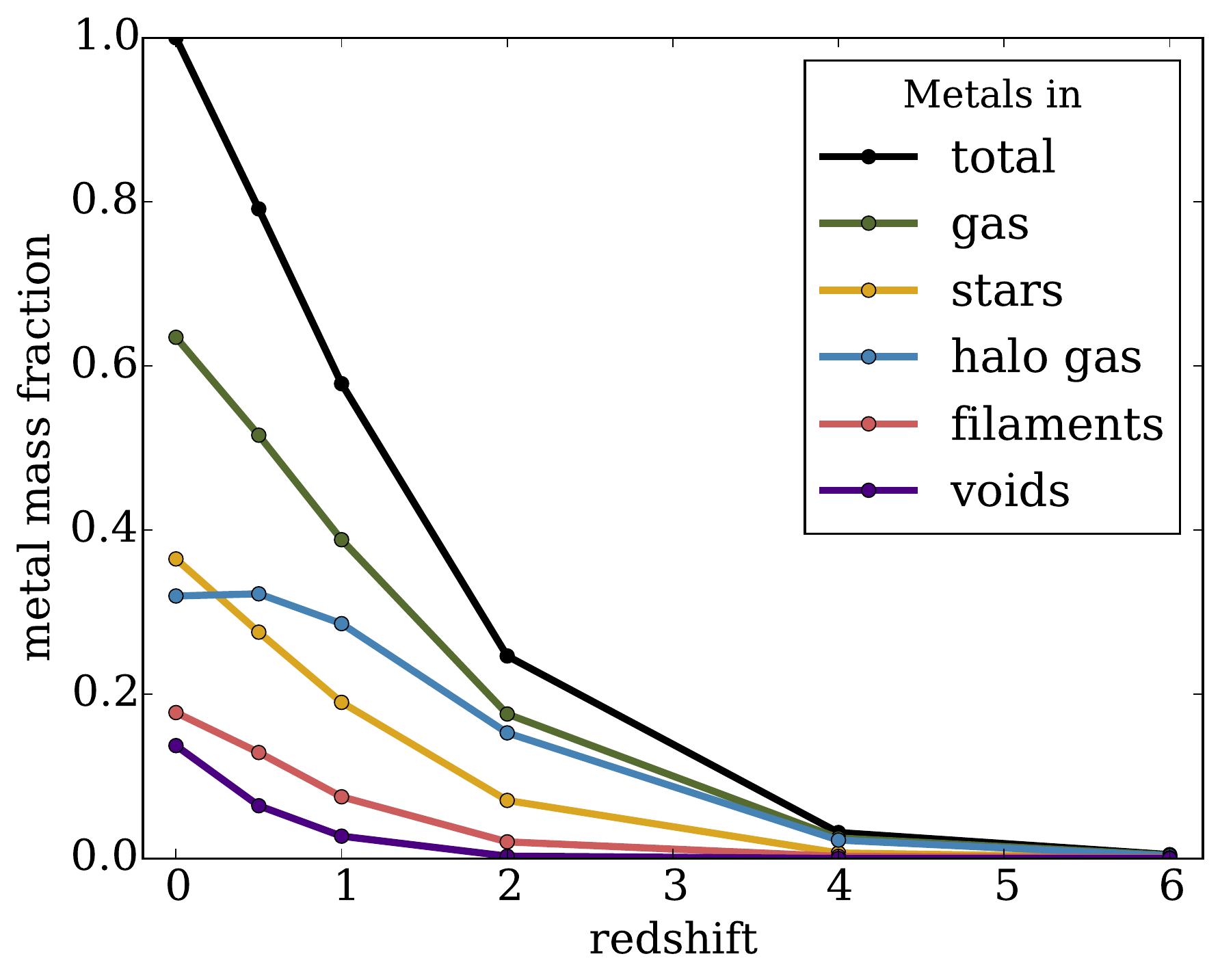}
\caption{Evolution of metal fractions in gas and stars. The values are normalized to the total metals at 
$z=0$, and the contribution of haloes, filaments and voids to the metals budget in gas is shown.}
\label{fig:metal_evolution}
\end{figure}

Illustris can also be used to probe the metal content in gas and stars. Star particles are stochastically 
formed when gas reaches densities above a threshold value $\rho_\textrm{sfr}$. A star particle is modelled as 
a single-age stellar population, for which the mass and metal return will be computed at every time-step. 
This material is then distributed over neighbouring gas cells \citep[see Section 2.2 and 2.3 
in][]{Vogelsberger2013}. In \autoref{fig:metal_evolution}, we show the evolution of the metals in gas and 
stars normalized to the total metal mass at $z=0$. Additionally, the figure shows which fraction of the 
metals in gas is residing in haloes, filaments or voids. 
We find that at $z=0$, \mbox{36 \%} of the metals are locked up in stars and \mbox{64 \%} of the metals are 
in gas. Considering only the gaseous component, half the metals are within haloes and 28 \% reside in the 
filaments. The remaining 22 \% are located in voids. The average metallicity of the stars is 1.49 solar 
metallicities at $z=0$, while the halo gas has about 0.37 solar metallicities on average. The average 
metallicity in filaments and voids is roughly 0.1 times the solar value.

\subsection{Baryonic temperature--density relation}\label{sec:whim}

\begin{figure} \centering \includegraphics[width=0.53\textwidth]{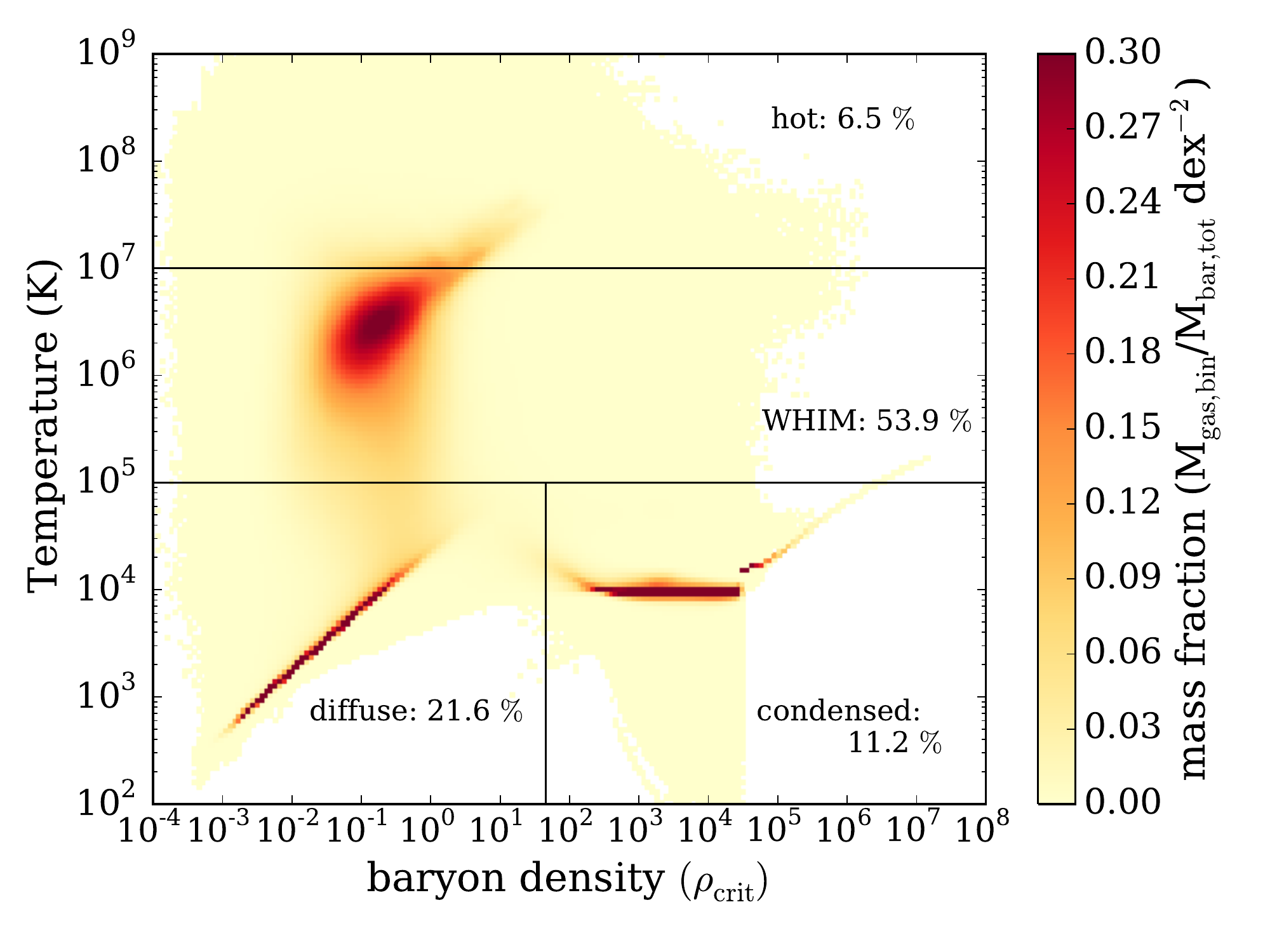} 
\caption{Mass fraction of gas to total baryons in temperature versus baryon density space. We subdivide the 
histogram into hot gas, diffuse gas, condensed gas and warm--hot intergalactic 
medium (WHIM). The mass fractions in these regions at $z=0$ are indicated in the figure.}  
\label{fig:whim_histo} \end{figure}

An alternative way to look at the distribution of baryons is to analyse them according to their density and 
temperature. This is more directly relevant to observations, as density and temperature are the important 
variables for emission and absorption mechanisms. In \autoref{fig:whim_histo}, we show the contribution of gas 
to the total baryonic mass in a temperature versus baryon density histogram. For this plot we directly use the 
Voronoi cell densities of the gas instead of the averaged densities we used in \autoref{sec:dm_density}. We 
divide the baryons according to the same classification as has been used in \cite{Dave2000}: diffuse 
gas having $\rho<1000$\rhocrit\OmegaB\ and $T<10^5$\,K, condensed gas with 
$\rho>1000$\rhocrit\OmegaB\ and $T<10^5$\,K, warm--hot gas with temperatures in the range 
$10^5<T<10^7$\,K and hot gas with temperatures above $10^7$\,K. We refer to the warm--hot gas also as 
WHIM.

We find that \mbox{21.6 \%} of the baryons are in the form of diffuse gas, located mainly in the 
intergalactic medium. The tight relation between temperature and density is due to the interplay of cooling 
through adiabatic expansion and photoionization heating (see also the discussion in Section 3.3 of 
Vogelsberger et al., 2012). The condensed gas amounts to \mbox{11.2 \%}.  Most of the condensed gas is in a 
horizontal stripe around 10$^4$ K. As photoionization heating becomes less dominant but cooling time-scales 
shorten at higher densities, gas cools effectively down to this temperature. Since there is no metal and 
molecular line cooling, 10$^4$ K represents an effective cooling floor for the gas in this phase. The upward 
rising slope which extends from the `condensed' region into the `WHIM' region is due to an effective equation 
of state used for gas exceeding the density threshold for star formation \citep{Springel2003}.  The warm--hot 
medium makes up for \mbox{53.9 \%} of the baryons, while \mbox{6.5 \%} 
of the baryons 
are in hot gas.

The mass in these two categories corresponds to warm--hot shock-heated gas in haloes and filaments and 
material which has been ejected due to feedback from haloes (inspection of \autoref{fig:temperature_slice} 
shows that the ejected material is in the temperature range defining the WHIM region). If all the \mbox{23.6 
\%} of the ejected material (see \autoref{table:regions}) had remained in haloes, this would have changed the 
warm--hot mass fraction to \mbox{30 \%} and increased the condensed fraction to 
\mbox{34.8 \%} (plus \mbox{6.6 \%} in stars).

\begin{figure} \centering 
\includegraphics[width=0.48\textwidth]{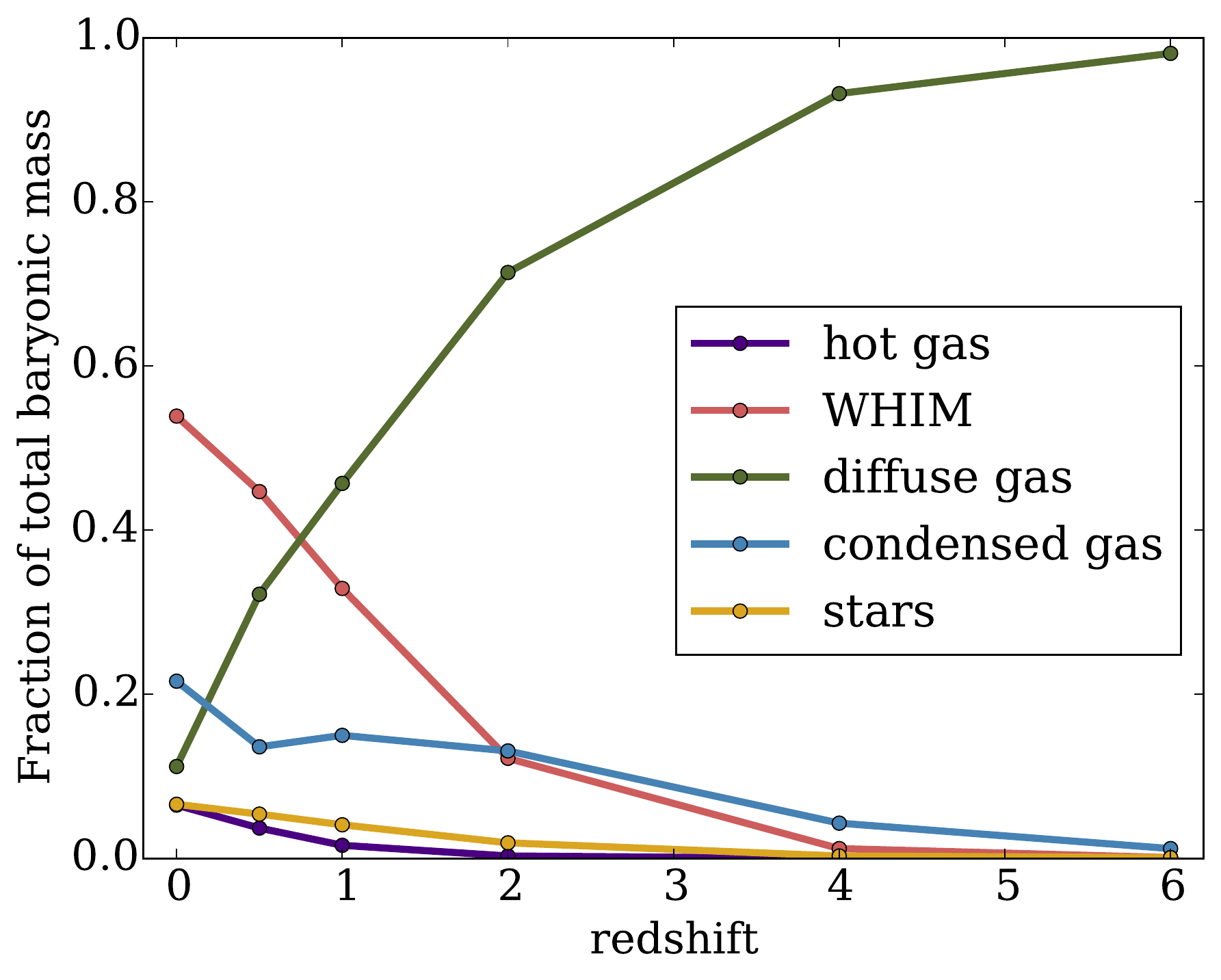} \caption{Redshift evolution of the mass 
fraction in the gas phases of \autoref{fig:whim_histo}. In addition to the gas phases, we also include the 
contribution of stars.}  \label{fig:whim_evo} \end{figure}

The redshift evolution of the gas phases is given in \autoref{fig:whim_evo}, where we see that at high 
redshift, most of the gas has been in the form of a rather cold and diffuse medium. Starting at a redshift of 
$z=4$, and more pronounced after redshift $z=2$, the WHIM phase is gaining more and more mass, and by 
redshift zero, ends up containing most of the baryons.  

 

\section{Discussion}\label{sec:discussion}
Comparing the values of \autoref{table:mass_in_halos} with \autoref{table:regions}, we see a very good 
agreement at $z=0$ between the \textsc{subfind} halo catalogue and the haloes defined by a dark matter 
density cut. This suggests that our method of measuring the mass using the average dark matter density in a 
cell of our grid works reasonably well. However, we see deviations of \mbox{5--7 \%} for the mass in haloes at 
redshifts higher than 1. The reason for these deviations is that at higher redshifts the haloes are less 
massive, and thus some fall below the resolution limit of our grid.

Using the temperature-baryon density classification in \autoref{sec:whim}, we find that \mbox{53.9 \%} of the 
baryons reside in the WHIM region. This is higher than the 30--\mbox{40 \%} found in the work of 
\cite{Dave2000}, or the work of \cite{Cen2005}, who reported between 40 and \mbox{50 \%} in the WHIM phase. 
This discrepancy is most likely due to the use of different feedback models. The importance of the feedback 
model is further underlined by the differences between the full physics and 
the non-radiative runs, which produce a nearly identical dark matter distribution but very different baryon 
distributions (see \autoref{fig:mass_distro_z0}). In the full physics run only half the baryons are within 
haloes compared to the non-radiative simulation (see \autoref{table:mass_in_halos}). This difference is even 
more significant when we consider that the non-radiative run contains no radiative cooling and therefore the 
baryons cannot condense as easily into haloes as in the full physics run.

As stated in \autoref{sec:cosmic_web}, it is interesting to note that we find 30 \% of the baryons 
residing in voids. The reason for this is the strong radio-mode AGN feedback model, which 
is expelling baryonic matter from haloes and is responsible for nearly 80 \% of the baryons 
ending up in voids. In a recent survey, \cite{Shull2012a} find that 30 \% of the baryons are missing in the 
low-redshift Universe. It would be tempting to explain these missing baryons with the baryons we find in 
voids. However, we have to be cautious in reaching this conclusion, as it is not clear whether the radio-mode 
feedback model implemented in Illustris is reliable.  An indication that the feedback model of Illustris is 
too strong comes from the gas content of massive haloes. Observations of galaxy cluster sized haloes find 
baryon fractions between 70 and \mbox{100 \%} of the primordial value \citep{Vikhlinin2005}. In contrast, we 
find (see \autoref{table:mass_in_halos}) that the Illustris simulation produces a baryon fraction of only 
\mbox{50 \%} for
haloes more massive than $10^{14} M_\odot$. The low gas content of massive haloes in the simulation has been 
attributed to the strong AGN radio-mode feedback \citep{Genel2014}. The radio mode is used when the 
accretion rate on to a black hole is low, and thus is strongest at low redshifts. Therefore, our results at low 
redshifts would most likely change if a weaker AGN feedback model was used.

Do the missing baryons in massive haloes explain the baryons we find in voids? The 149 haloes with a total 
mass higher than 10$^{13}$M$_\odot$ contain 20.1 \% of the total dark matter but only 7.2 \% of the baryonic 
mass. Thus, 13 \% of the total baryons are missing in these haloes. It therefore seems that the missing mass 
cannot explain the full 24 \% of ejected material in voids. However, it is possible that material which 
is pushed out from the haloes might remove gas from the filaments. We should also note that the 24 \% we 
quote is only the ejected mass inside voids. The actually ejected mass might be higher as some of the ejected 
material would still be in the filaments.

Other simulations of structure formation which include AGN
  feedback have produced baryon fractions in massive haloes in
  agreement with observations [\citet{McCarthy2010},
  \citet{LeBrun2014}, \citet{Schaye2014}, \citet{Planelles2013} or
  \citet{Khandai2014}]. It is beyond the scope of this paper to give a
  detailed comparison of different approaches. However, as the AGN
  feedback has a large impact on our results, we want to highlight some
  differences in the models. Most current cosmological simulations,
  including Illustris, include AGN feedback through modifications and
  extensions of the approach introduced in
  \citet{Springel2005b}. There, black holes are seeded when the
  hosting halo passes a threshold mass
  ($5\times 10^{10}\,h^{-1} \textrm{M}_\odot$ for Illustris), and then
  Eddington-limited Bondi--Hoyle accretion is assumed. As the actual
  accretion region cannot be resolved, and the medium there is usually
  a mixture of hot gas and cold clouds, the Bondi--Hoyle accretion is
  sometimes enhanced by a factor $\alpha$, which is invoked to
  compensate for the lack of resolution. In Illustris, this accretion
  enhancement parameter is set to $\alpha=100$. It is further assumed
  that a fraction $\epsilon_\textrm{\small r}$ (set to 0.2 in
  Illustris) of the rest mass energy of the accreted mass is converted
  into energy, and a fraction of this energy in turn couples to the
  surrounding material. In Illustris, a modification of the model by
  \citet{Sijacki2007} is used, where a distinction between a
  quasar mode and a radio mode has been introduced. The
  quasar mode is assumed to be active if the accretion rate is above
  \mbox{5 \%} of the Eddington rate, and the radio mode, is used below
  this threshold.  In the quasar mode, a fraction
  $\epsilon_\textrm{\small f}=0.05$ of the AGN luminosity is coupled
  thermally to cells around the black hole. In the radio mode, the AGN
  is assumed to inject energy into bubbles, whose radius and distance
  to the black hole are computed according to \citet{Sijacki2007}. The
  radio mode is assumed to have a higher coupling efficiency of
  $\epsilon_\textrm{\small m}=0.35$. The radio mode bubbles are not
  formed continuously, but only after the mass of the black hole
  increased by a specified value $\delta_\textrm{\small BH}$ (15 \% in
  Illustris). If the fractional mass increase $\Delta M/M$ surpasses
  $\delta_\textrm{\small BH}$, the energy
  $\epsilon_\textrm{\small m}\epsilon_\textrm{\small r}\Delta M c^2$
  is put into a bubble. Thus, $\delta_\textrm{\small BH}$ controls the
  frequency and energy of bubbles.

  Among the recent hydrodynamical simulations of galaxy formation 
  mentioned above, Illustris is the only simulation which makes a
  distinction between radio and quasar mode in this way. \citet{Planelles2013}
  also use different efficiencies for radio and quasar mode, but the
  energy is coupled to surrounding particles in the same way for both
  modes. The model of \citet{Booth2009}, which is used in
  \citet{McCarthy2010a}, \citet{LeBrun2014} and \citet{Schaye2014}
  does not discern between radio and quasar mode. A further difference
  between Illustris and \citet{Booth2009} is that they use a different accretion model. Instead of
  a constant value for the accretion enhancement, $\alpha$ is set to 1 at low gas densities 
  and varies as a power law of density at higher densities \citep[see][for a
  discussion]{Booth2009}. In addition to this, \citet{Schaye2014} 
  include a model which can lower the accretion rate in order to account for the 
  angular momentum of accreting gas \citep[see][]{Rosas-Guevara2015}.
  Also, they use a radiation efficiency of $\epsilon_\textrm{\small r}=0.1$
  and a coupling efficiency of 
  $\epsilon_\textrm{\small f} = 0.15$. In order to prevent that the
  injected energy is cooled away immediately, the feedback energy is
  not emitted continuously, but only once enough energy has
  accumulated to lead to a heating $\Delta T$ of $10^8$ K. The energy
  in then injected into one particle close to the black hole (or
  multiple particles if the energy produced per time-step is high
  enough to heat several particles to $10^8\,\textrm{K}$).

  There are also differences with respect to the used hydrodynamical
  scheme.
 \citet{Vogelsberger2011}
  found that the SPH code \textsc{gadget3} produces higher
  temperatures in haloes than the moving-mesh code \textsc{arepo},
  which has been employed in Illustris. They argue that this higher
  temperature has to do with spurious dissipation in SPH related to
  the higher numerical noise in this method. This might partly explain
  why the AGN feedback efficiency parameters had to be set higher in
  Illustris than in a comparable SPH simulation in order to reach the
  same suppression of star formation in massive haloes.
 Also, the mass resolution in \citet{McCarthy2010a} and
  \citet{LeBrun2014} is very low compared to Illustris, which could
  lead to different density estimates at the black hole and thus
  influence the accretion rates significantly, even if the same
  accretion model was used.
  
For lower mass haloes AGN feedback is less important, as lower mass haloes typically
have less massive black holes and the AGN luminosity scales with the square of black 
hole mass. Therefore, at low to medium masses the dominant feedback process regulating 
star formation is stellar feedback. A part of the stellar feedback is already implicitly 
included through the effective equation of state for star forming gas \citep{Springel2003}.
In addition to this, Illustris uses a non-local `wind' feedback scheme (see \citet{Vogelsberger2013}), where a 
part of the star forming gas can be stochastically converted into wind particles. These wind
particles are allowed to travel freely for some time (or until a density threshold is reached) before their
mass and energy is deposited into the appropriate gas cell. Without these feedback processes, galaxies
would form too many stars (the so called over-ooling problem). Thus the parameters of this models will
also influence the results we presented (especially the region below $10^{12}$ \Msol in
\autoref{fig:baryon_fraction_halos}). The Illustris stellar mass function is a bit too high in this mass
range, suggesting that the feedback should be more effective in preventing star formation. Similarly, we find 
a high baryon fraction around unity in these haloes. As stated in \autoref{sec:results}, there are substantial
observational uncertainties for the baryon fraction in this mass range, and the observations we included in
\autoref{fig:baryon_fraction_halos} are to be thought of as lower limits.

We should caution however that while the feedback of Illustris is so strong that it depletes the massive 
haloes of their gas, the stellar mass function of Illustris is still too high for haloes with a stellar mass above $10^{12} M_\odot$ \citep[see Fig.~3 
in][]{Genel2014}. This indicates that the feedback is not efficient enough in suppressing star formation at 
this mass scale. Simply tuning the energy parameters of the feedback model cannot resolve the tension 
between the low gas content in massive haloes on one hand and the too high stellar mass function on the other 
hand. Either the AGN feedback model has to be modified \citep[][suggests making the duty cycle longer and 
thus avoiding short strong bursts]{Genel2014} or this is a hint that other, so far neglected processes are 
important. It will be interesting to see if improved feedback models, which reproduce the baryon fraction in 
massive haloes correctly, will continue to find such a 
significant amount of baryons in dark matter voids.

\section{Summary}\label{sec:summary}

We investigated the global mass distribution of baryons and dark matter in different structural components of 
the Illustris simulation. By attributing dark matter density regions to the constituents of the cosmic web, 
we could measure the mass and volume of haloes, filaments and voids, as summarized in \autoref{table:regions}.

At redshift $z=0$ we find that the haloes host \mbox{44.9 \%} of the total mass, \mbox{49.2 \%} of the dark 
matter and  \mbox{23.2 \%} of the baryons. Their volume fraction amounts to only  \mbox{0.16 \%} of the total 
simulation volume. The filaments host \mbox{44.5 \%} of the dark matter and \mbox{46.4 \%} of the baryons and 
correspond to \mbox{21.6 \%} of the simulation volume. The dark matter voids contain \mbox{6.4 \%} of the 
dark matter but \mbox{30.4 \%} of the baryons, and the volume fraction of the voids amounts to \mbox{78.2 
\%}. About \mbox{80 \%} of the baryons inside voids correspond to gas which has been ejected from haloes due 
to feedback processes. Most of this material, which occupies one third of the volume of the voids, has been 
ejected at redshifts lower than $z = 1$.

An analysis of a comparison run, in which star formation, cooling and feedback had been switched off, showed 
that the baryons trace dark matter well in a non-radiative model, which is not the case for the full physics 
simulation. This demonstrates that the feedback model has significant consequences for the global mass 
distribution. While the feedback model of Illustris succeeds in reproducing a realistic stellar mass 
function, the low baryon fraction in massive haloes suggests that the feedback model is not correctly 
reproducing their gas content. At the moment it is still unclear whether our analysis would produce similar 
results with a different feedback model. Work on improved feedback models is ongoing, and it will be 
interesting to clarify this question in future studies based on a next generation of simulation models.

\section*{Acknowledgements}
MH and DS thank their colleagues at the Institute for Astro- and Particle Physics at the University of 
Innsbruck for useful discussions, especially Francine Marleau, Dominic Clancy, Rebecca Habas and Matteo 
Bianconi. DS acknowledges the research grant from the office of the vice rector for research of the 
University of Innsbruck (project DB: 194272) and the doctoral school - Computational Interdisciplinary 
Modelling FWF DK-plus (W1227). This work was supported by the Austrian Federal Ministry of Science, Research 
and Economy as part of the UniInfrastrukturprogramm of the Focal Point Scientific Computing at the University 
of Innsbruck. SG acknowledges support provided by NASA through Hubble Fellowship grant HST-HF2-51341.001-A 
awarded by the STScI, which is operated by the Association of Universities for Research in Astronomy, Inc., 
for NASA, under contract NAS5-26555. VS acknowledges support through the European Research Council through 
ERC-StG grant EXAGAL-308037.

\bibliographystyle{mnras}

\label{lastpage}

\end{document}